\documentclass[%
 reprint,
preprintnumbers,
 amsmath,amssymb,
 aps,
]{revtex4}

\usepackage{graphicx}% Include figure files
\usepackage{dcolumn}% Align table columns on decimal point
\usepackage{bm}% bold math
\usepackage{hyperref}% add hypertext capabilities
\usepackage{braket}
\usepackage{physics}
\usepackage{qcircuit}
\usepackage{float}
\usepackage{graphicx}
\usepackage{feynmp-auto}
\usepackage{wrapfig}
\usepackage{longtable}
\usepackage[caption=false]{subfig}

\begin{document}
\preprint{INT-PUB-20-027}
\title{An Algorithm for Quantum Computation of Particle Decays}

\author{Anthony Ciavarella}
\email{aciavare@uw.edu}
\affiliation{Institute for Nuclear Theory, University of Washington, Seattle, WA 98195-1550, USA}

\date{\today}

\begin{abstract}
A quantum algorithm is developed to calculate decay rates and cross sections using quantum resources that scale polynomially in the system size assuming similar scaling for state preparation and time evolution. This is done by computing finite-volume one- and two-particle Green's functions on the quantum hardware. Particle decay rates and two particle scattering cross sections are extracted from the imaginary parts of the Green's function. A $0+1$ dimensional implementation of this method is demonstrated on IBM's superconducting quantum hardware for the decay of a heavy scalar particle to a pair of light scalars.
\end{abstract}

\maketitle

%\tableofcontents

\section{Introduction}
Quantum field theories describe three of the four fundamental forces of nature. In particular, Quantum chromodynamics (QCD) describes the strong interactions that bind together quarks into hadrons \cite{Schwartz}. The predictions of QCD are often tested in experiments where unstable hadrons decay and their decay products are observed. For high energy phenomenon, Feynman diagrams and other perturbative techniques provide an excellent description. In the low energy region, the QCD coupling constant becomes large and these methods fail. Nonperturbative approaches such as lattice QCD (LQCD), chirial perturbation theory and other effective field theories have enabled the calculation of some hadronic properties in this region. For example, Luscher's method \cite{Luscher:1985dn,Luscher:1986pf} has allowed the computation of some decay rates and scattering cross sections using LQCD by relating them to finite volume energy shifts. It has been used to compute scattering phase shifts for several low energy processes \cite{NPL1,NPL2,NPL3,NPL4}, and the decay widths of $\rho$ and $\sigma$ mesons \cite{rhomeson,sigmameson}. The extraction of finite volume energy levels becomes difficult for excited states and for large lattices which limits the applicability of the method. 

Quantum computers have been proposed as a tool to avoid various problems present in simulations of QFT's. In particular, fault tolerant quantum computers are expected to be capable of simulating time evolution of local QFT's using resources that scale polynomially in the system size \cite{Preskill1,Preskill2,Preskill3}. The first steps towards simulating lattice gauge theories, such as the Schwinger Model, have been made \cite{Klco:2018kyo,Schwinger1,Schwinger2,Schwinger3,Schwinger4,Schwinger5,Schwinger6,LGT1,LGT2,su2sim,Zohar_2013,Zohar_2015,Lamm_2019,Alexandru_2019,Alexandru_2019_2,Lamm_2020,Ba_uls_2020,Tagliacozzo_2013,Tagliacozzo_2013_2}. In this work, a method of extracting particle decay rates and scattering cross sections from a Green's function calculated on a quantum computer is demonstrated. This method only requires the ability to prepare initial particle states and perform real time evolution. It has been shown for scalar and fermionic field theories that state preparation and real time evolution can be performed on quantum computers using resources that scale polynomially with the system size \cite{Preskill1,Preskill2,Preskill3}. The computational costs of classically performing real time evolution usually scales exponentially with the system size \cite{Sign1,Sign2,Sign3,Sign4,Sign5,Sign6,Sign7,Sign8} so the use of quantum computers would represent an exponential speedup. A classical simulation of this quantum algorithm is explicitly demonstrated for a 1+1 dimensional QFT where a heavy scalar decays to a pair of light scalars. A $0+1$ dimensional demonstration is performed using IBM's superconducting hardware. Although this calculation is demonstrated for a specific model, the approach is based on general properties of Green's functions and it is expected that it can be applied to particle decays or scattering in other theories.

The paper is organized as follows. The method of computing the decay rate from the Green's function is described in Section \ref{sec:GF} and the mathematical details are shown in Appendix \ref{appendix:poles}. The quantum circuit used to calculate the Green's function is described in Appendix \ref{appendix:quantumgreen}. The time truncation and discretization errors are analyzed in Appendix \ref{appendix:Terrors}. The systematic errors present in extracting a decay rate from a finite volume Green's function are analyzed in Appendices \ref{appendix:FVErr} and \ref{appendix:epserror}. The errors due to finite particle number truncations for theories containing bosons are analyzed in Appendix \ref{appendix:ParticleTruncation}. A classical simulation of this quantum algorithm is performed in Section \ref{classicalsim}. IBM's quantum processor is used to implement this algorithm in Section \ref{ibm}. The Trotterization procedure used in this demonstration is described in Appendix \ref{appendix:hamsim}. The data from running on IBM's quantum processor is in Appendix \ref{appendix:data}.

\section{Quantum Computation of Green's Functions}
\label{sec:GF}
For a single particle state $\ket{\psi}$, the Green's function can be written as $\bra{\psi} \frac{1}{\omega - \hat{H} + i \eta} \ket{\psi} = \frac{1}{\omega - E + i\eta - \bra{\psi} \hat{T}(\omega + i \eta) \ket{\psi}}$, where $E$ is the energy of the state $\ket{\psi}$, $\hat{H}$ is the Hamiltonian and $\hat{T}$ is the scattering $T$ matrix as shown in Appendix \ref{appendix:poles}. If the Hamiltonian, $H$, can be split into a free piece $H_0$ that describes the propagation of free particles and an interaction piece $V$ that describes the interaction of particles, the state $\ket{\psi}$ can be prepared on a quantum computer as an eigenstate of $H_0$ using previously developed methods \cite{Preskill1,Preskill2,tensornetwork}. For theories like QCD, where no such division is known, an unstable particle state can be prepared by simulating two stable particles colliding on resonance. For example, a $\rho$ meson can be prepared by simulating the collision of two pions with total energy equal to the $\rho$ meson mass. The inclusive decay rate of a particle in $d$ spatial dimensions is given by
\begin{equation}
\Gamma = \sum_{X_f} \int d P_{X_f} (2 \pi)^{d+1} \delta^{d+1} (P_{X_f} - P_{\psi}) \abs{\bra{X_f} \hat{T}(E_{\psi}) \ket{\psi}}^2
\end{equation} 
where $P_{\psi}$ is the energy-momentum vector of the initial particle, $P_{X_f}$ is the energy-momentum vector of the final state $X_f$, the sum is performed over all possible final states and the integral is performed over all possible energy-momenta vectors of the final state. The optical theorem relates this sum to the forward matrix element of the $T$ matrix by $\Gamma = -2 \lim\limits_{\eta \rightarrow 0} \Im(\bra{\psi} \hat{T}(E + i \eta) \ket{\psi})$ \cite{Schwartz}. Therefore, if the Green's function can be computed in the $\eta \rightarrow 0$ limit, the inclusive decay rate can be extracted from it. For $\eta \neq 0$, the difference between $\Im(\bra{\psi} \hat{T}(E + i \eta) \ket{\psi}$ and $\Gamma$ is $O(\eta)$ as shown in Appendix \ref{appendix:epserror}. Furthermore, if $\ket{\psi}$ is a two particle state, the same kind of relationship between the Green's function and the $T$ matrix holds, and the optical theorem can be used to find the inclusive scattering cross section for the two particles present in the state. To simplify the following discussion, the case of particle decays will be focused on. When the theory describing the particle is simulated inside a finite volume box with periodic boundary conditions, the difference between $\Im(\bra{\psi} \hat{T}(E + i \eta)\ket{\psi})$ in the finite volume and the infinite volume value for a $1 \rightarrow N$ decay is $O(E^{\frac{d-1}{2}N-2} e^{-\frac{\eta}{N+1} \frac{L}{2}})$ for a $d+1$ dimensional theory with a mass gap, and  $O(\frac{1}{\eta^2 L})$ otherwise, where $L$ is the length of a side of the finite volume box, as shown in Appendix \ref{appendix:FVErr}. Therefore, if the Green's function can be calculated in a finite volume for a theory with a mass gap, $\Gamma$ for $1 \rightarrow N$ decays can be determined with finite $\eta$ errors that are $O(\eta)$ and finite volume errors that are $O(M^{\frac{d-1}{2}N-2} e^{-\frac{\eta}{N+1} \frac{L}{2}})$. It should be noted that the $L \rightarrow \infty$ and $\eta \rightarrow 0$ limits are not independent and to have finite volume errors vanish in the $L \rightarrow \infty$ limit, $\eta$ must be chosen such that $\eta L \rightarrow \infty$. To evaluate this Green's function, it is helpful to express it in integral form,
\begin{equation}
\bra{\psi} \frac{1}{\omega - \hat{H} + i \eta} \ket{\psi} = - i \int_{0}^{\infty} \bra{\psi} e^{i(\omega + i \eta - \hat{H})t} \ket{\psi} dt \ \ \ .
\end{equation}
If this integral is truncated at finite time $T$, a Riemann sum approximation,
\begin{equation}
R = \sum_{k=0}^{T/\Delta t} e^{i(\omega + i \eta) k \Delta t} \bra{\psi} e^{-i \hat{H} k \Delta t} \ket{\psi} \Delta t \ \ \ ,
\end{equation}
to this integral can be evaluated on a quantum computer with the techniques described in Appendix \ref{appendix:quantumgreen} within an accuracy of $\epsilon$ using a gate count that scales as 
\begin{equation}
\text{Gate Count} = O\left(\frac{\log(\frac{2}{\epsilon \eta})}{\eta^2 \epsilon} \ (\omega + E + \eta) \ p\left(\frac{1}{\eta} \log(\frac{2}{\eta \epsilon}),\eta \epsilon \right) \right)
\end{equation}
where $p(t, \delta)$ is the gate count required to evolve to time $t$ with accuracy $\delta$, provided that $\ket{\psi}$ has already been prepared and $E$ is the energy of the state $\ket{\psi}$. Once the Green's function has been computed, the particle decay rate can be extracted from the imaginary part of its pole. It should be noted that in the $\eta \rightarrow 0 $ limit, the imaginary part of the Green's function becomes the spectral density function and other work has been done on using quantum computers to calculate the spectral density function \cite{spectral1,spectral2,spectral3}. $\Gamma$ can be extracted from the peak of the Green's function which takes the value $\frac{2}{\Gamma}$. Therefore, to compute $\Gamma$ to within an accuracy $\delta \Gamma$, the Green's function must be computed to within an accuracy of $\frac{\delta\Gamma}{\Gamma^2}$. Since the uncertainty in $\Gamma$ scales linearly with $\eta$, $\Gamma$ can be determined to an accuracy of $\delta \Gamma$ using 
\begin{equation}
\text{Gate Count} = O\left(\frac{\log(\frac{2\delta \Gamma}{\Gamma})}{\delta \Gamma^3} \ \Gamma^2 \ (2E + \delta \Gamma) \ p \left(\frac{2}{\delta \Gamma} \log(\frac{\sqrt{2}\Gamma}{\delta \Gamma}),\left(\frac{\delta \Gamma}{\Gamma}\right)^2 \right)\right)
\end{equation}
gates with a lattice whose size scales as $O(\frac{1}{\delta \Gamma} \log(\frac{1}{\delta \Gamma}))$ when the theory has a mass gap.

Another approach to computing the decay rate of an unstable particle would be to prepare the initial state, evolve for some time and measure detector operators at the border of the box, similar to the algorithm for the scattering of scalar particles in previous work \cite{Preskill1}. This requires the simulation to run for a time $t = O(\frac{1}{\Gamma})$ before measuring the detector operators. The algorithm presented here only requires the simulation to run for a time $t = \frac{2}{\delta \Gamma} \log(\frac{\sqrt{2}\Gamma}{\delta \Gamma})$. Therefore, this algorithm is expected to perform better for particles with a long lifetime. This algorithm also provides a method of computing decay rates that is different from direct time evolution and should have different systematic errors. Comparing decay rates computed with these two different methods will allow them to be determined with a higher degree of confidence.

\section{Decay of a Heavy Scalar}
\label{classicalsim}
A demonstration of the algorithm discussed in previous sections will be provided by a classical simulation of the decay of a heavy scalar, $\phi$, to a pair of light scalars, $\chi$, in 1+1 dimensions. The Lagrangian for this process is given by
\begin{equation}
\mathcal{L} = \frac{1}{2}(\partial \phi)^2 + \frac{1}{2}(\partial \chi)^2 - \frac{1}{2}M_0^2 \phi^2 - \frac{1}{2} m_0^2\chi^2 - \frac{1}{2} g \phi \chi^2 - \frac{1}{4!} \lambda \chi^4
\end{equation}
where $M_0$ and $m_0$ have been chosen such that the heavy particle's mass is $2.01$ times the light particle's mass (so the  $\phi \rightarrow 2 \chi$ channel is the only allowed decay channel) and $\lambda > \frac{3 g}{M_0^2}$ (to ensure a stable vacuum without spontaneous symmetry breaking in the infinite volume continuum theory). This theory was placed on a lattice with periodic boundary conditions and with lattice spacing $a = 0.2 m^{-1}$ where $m$ is the light particle's mass. This was done for lattices with three, five and seven sites. With these boundary conditions, the allowed momentum modes are in the set
$\{-\frac{\pi (n_s - 1) }{L}, -\frac{ \pi (n_s - 1) }{L} + \frac{2 \pi}{L},..., \frac{\pi (n_s - 1) }{L} \} $, where $n_s$ is the number of sites and $L = n_s a$ is the length of the finite volume box. To simulate this on a classical computer, the $\phi$ occupation numbers were truncated at one for each momentum mode and the $\chi$ occupation numbers were truncated at two for each momentum mode. Since the mass of the heavy particle is only slightly larger than two times the light particle's mass, the arguments of Appendix \ref{appendix:ParticleTruncation} indicate that the error in the decay rate calculation due to this particle number truncation should be negligible.

A classical computer was used to determine the renormalization parameters, and to simulate the quantum algorithm from the previous section. The renormalization conditions were that the vacuum has zero energy and the mass of the heavy scalar is 2.01 times the mass of the light scalar. For each lattice volume, $\eta$ was chosen to minimize the sum of the finite volume and finite $\eta$ error calculated using the methods in Appendices \ref{appendix:FVErr} and \ref{appendix:epserror}. 

\begin{figure}[H]
\centering
\includegraphics[height=4cm]{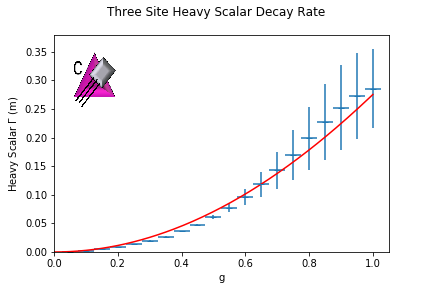}
\includegraphics[height=4cm]{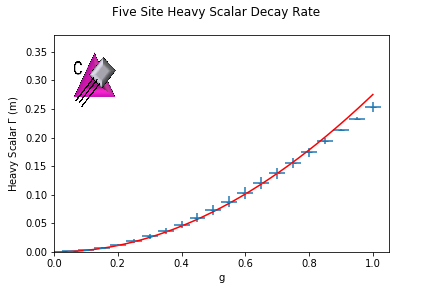}

\includegraphics[height=4cm]{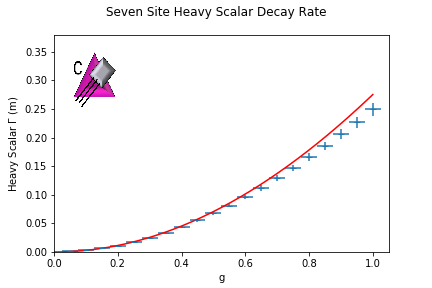}

\caption{Heavy particle decay rates calculated on different lattice volumes plotted as a function of the coupling constant. The blue points are the decay rates calculated in the classical simulations of the quantum algorithm and the red curves are the one loop infinite volume continuum calculation. The error bars on the finite lattice decay rates represent finite volume and finite $\eta$ errors calculated using the methods in Appendix \ref{appendix:errors}. The icons appearing are defined in Ref. \cite{Icons}.}
\label{fig:LatticeG}
\end{figure}

The heavy particle decay rates calculated classically in this example are displayed in Fig. \ref{fig:LatticeG}. The finite volume and finite $\eta$ uncertainties were calculated using the methods described in Appendix \ref{appendix:errors}. To improve the precision of this calculation, a larger lattice must eventually be used. No matter what truncation is used, the dimension of the Hilbert space will grow exponentially with the number of lattice sites. The Green's function can be computed on a classical computer using matrix inversion techniques, the fastest of which scale as the dimension of the Hilbert space which grows exponentially with the number of sites \cite{LITMethod,MatrixInversion}. Due to this exponential scaling, it is infeasible to use a classical computer to compute Green's functions on a large lattice. However, using previously developed techniques for simulating scalar field theories, the method described in the previous section can be used to compute the Green's function on a quantum computer using resources that scale polynomially \cite{Preskill1,ScalarBenchmark}.

\section{Demonstration of 0+1 Theory on IBM's Quantum Processor}
\label{ibm}
The calculations in the previous section were performed using classical computers, but it is possible to use existing quantum computers to do these calculations for a single lattice site with the truncations from the previous section. The Ourense quantum processor made available by IBM was used to implement this method for a one site calculation of the heavy particle decay rate.

\begin{figure}[H]
\centering
\includegraphics[height=5.5cm]{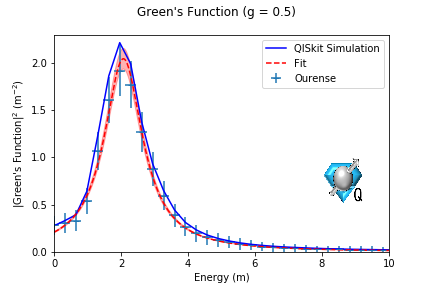}
\includegraphics[height=5.5cm]{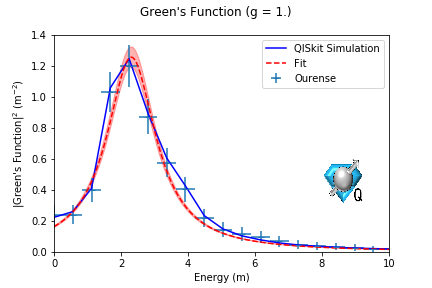}
\label{fig:width}
\caption{Green's functions computed with the Ourense quantum processor. The solid blue curve is a zero noise classical simulation of this calculation with Qiskit. The light blue points were computed using the error mitigated amplitudes from the Ourense quantum processor. The error bars represent uncertainties from the error mitigation extrapolation. The red curve is the Lorentzian fit to the error mitigated Green's functions.}
\end{figure}

The details of how the theory was discretized and how time evolution was implemented on the quantum computer are described in Appendix \ref{appendix:hamsim}. The Hadamard test method \cite{HadTest} was used to obtain $\bra{ \phi}e^{-i \hat{H} \Delta t k} \ket{\phi} $ for $\Delta t = 0.2 m^{-1}$ and $k=1,2,...,96$ , where $\ket{\phi}$ is a state describing a single heavy scalar at rest. Two Trotter steps were used to calculate each time slice so the circuits used to calculate the real component of $\bra{ \phi}e^{-i \hat{H} \Delta t k} \ket{\phi} $ used $36$ single qubit gates and $28$ CNOT gates. The circuit used to estimate the imaginary component had one additional single qubit gate. Due to the length of the circuit used, the effect of imperfect gate implementation on the Ourense quantum processor is non-negligible. The contribution of imperfect gate implementation to the error in the computed amplitudes was estimated using the technique described in Appendix \ref{appendix:gateerr}. Each circuit used in the Hadamard test was sampled 8000 times so the resulting statistical error was negligible relative to the systematic gate errors. To mitigate the effects of gate errors, an error mitigation technique described in Appendix \ref{appendix:mitigation} was used to extrapolate to the zero CNOT gate error limit. The Green's function, 
\begin{equation}
G = \abs{ \sum_k e^{i (\omega + i \eta)k \Delta t}\bra{\phi}e^{-i \hat{H} \Delta t k} \ket{\phi} \Delta t }^2 \ \ \ ,
\end{equation}
was calculated classically using the error mitigated amplitudes and the results for two different couplings are displayed in Fig. \ref{fig:width}. The heavy particle decay rate was extracted from the Green's function by performing a least squares fit to a Lorentzian distribution. The extracted decay rate is compared to the ideal decay rate,
\begin{equation}
\Gamma = -\Im\left(\bra{\phi} \hat{T}(M+i\eta)\ket{\phi}\right)
= \sum_n \frac{2 \eta}{(M-E_n)^2 + \eta^2} \abs{\bra{E_n} \hat{V} \ket{\phi} }^2 \ \ \ 
\end{equation}
where the states $\ket{E_n}$ are eigenstates of the Hamiltonian with energy $E_n$, that would be computed in the absence of any finite $T$ or $\Delta t$ errors in Table \ref{tab:results}.

\begin{table}[H]
\centering
\begin{tabular}{|l|l|l|}
\hline
g & Ideal $\Gamma$ & Extracted $\Gamma$ \\
\hline
0.5 & $0.070 m$ & $(0.099 \pm  0.037) m$ \\
1. & $0.287 m$  & $(0.286 \pm 0.047) m$ \\
\hline
\end{tabular}
\caption{Heavy particle decay rates calculated with the Ourense quantum processor. The first column is the coupling constant. The second column is the value of $\Gamma$ that would be computed in the absence of any finite $T$ or $\Delta t$ errors. The third column is the decay rate calculated with the Ourense quantum processor. The error represents uncertainties in the fit to the Green's function.}
\label{tab:results}
\end{table}
The heavy particle decay rates calculated on the Ourense quantum processor are in agreement with the ideal calculation. However, even after using these error mitigation techniques, the error due to imperfect gates remained large. 

\section{Conclusion}
In this work, a quantum algorithm to calculate the decay rate of unstable particles and scattering cross sections has been introduced. The resources required to implement this method scale polynomially with the system size provided that state preparation and time evolution can be performed using resources that scale polynomially in the system size and field value truncations. It has been shown that this is possible for scalar and fermionic field theories \cite{Preskill1,Preskill2,tensornetwork}. To apply this method to LQCD, it will be necessary to develop techniques to prepare hadronic states and perform time evolution in lattice gauge theories. IBM's Ourense quantum processor was used to apply this algorithm to a scalar field theory defined on a single lattice site with truncated occupation numbers. Bounds on the finite volume error of $1 \rightarrow N$ decay rates and $2 \rightarrow N$ scattering cross sections computed with this method have been determined. More work will need to be done to understand how different truncations effect the error in the computed decay rate. The method presented here only requires preparation of the initial state and the ability to simulate the Hamiltonian. Classical methods of computing decay rates and cross sections from lattice calculations such as Luscher's method rely on relating these observables to finite volume energy shifts. In general this is a difficult process, and only allows the calculation of decay rates and cross sections for limited processes. Due to the greater generality of this method, it is expected that quantum computers will be able to calculate decay rates and cross sections beyond the reach of classical computers.

\section{Acknowledgments}
We would like to thank Martin Savage for many useful discussions. We would like to thank Natalie Klco and Alessandro Roggero for feedback in preparing this manuscript. We thank the Institute for Nuclear Theory at the University of Washington for its kind hospitality and stimulating research environment. This research was supported in part by the INT's U.S. Department of Energy grant No. DE-FG02- 00ER41132. This work was supported
in part by the U. S. Department of Energy grant No. DE-SC0019478. We acknowledge use of the IBM Q experience for this work. This work was facilitated through the use of advanced computational, storage, and networking infrastructure provided by the Hyak supercomputer system at the University of Washington.

\nocite{Icons}

\appendix

\section{Green's Function Poles}
\label{appendix:poles}
The Green's function used in this method is 
\begin{equation}
G = \bra{\psi} \frac{1}{\omega - \hat{H} + i \eta} \ket{\psi} \ \ \ ,
\end{equation}
where $\ket{\psi}$ is a state describing the particle that will be decaying and $\hat{H}$ is the Hamiltonian of the system. This Green's function has poles whose real part is the energy of the state $\ket{\psi}$ and whose imaginary part is given by the imaginary part of the forward scattering amplitude. The manipulations to show this are standard \cite{Luscher:1986pf}, but have been reproduced here for the reader's convenience. The Hamiltonian can be split into a free term and an interaction term so that $\hat{H} = \hat{H}_0 + \hat{V}$, $\hat{H}_0 \ket{\psi} = E_0 \ket{\psi}$. Let $\hat{P} = \ket{\psi}\bra{\psi}$, $\hat{Q} = 1 - \ket{\psi}\bra{\psi}$. The Green's function can be written as
\begin{equation}
G = \frac{1}{\omega - E_0 + i \eta} + \frac{1}{(\omega - E_0 + i \eta)^2} \bra{\psi}\hat{V} \sum_{n=0}^{\infty}\left(\frac{1}{\omega - \hat{H}_0 + i \eta} \hat{V}\right)^n  \ket{\psi} \ \ \ .
\end{equation}
Using the matrix identity 
\begin{equation}
\hat{A} \sum_{n=0}^{\infty} ((\hat{B} + \hat{C}) \hat{A})^n = \hat{A}' \sum_{n=0}^{\infty} (\hat{B} \hat{A}')^n
\end{equation}
where $\hat{A}' = \hat{A} \sum \limits_{n=0}^{\infty} (\hat{C} \hat{A})^n$ with $\hat{A} = \hat{V}$, $\hat{B} = \frac{\hat{P}}{\omega - \hat{H}_0 + i \eta}$ and $\hat{C} = \frac{\hat{Q}}{\omega - \hat{H}_0 + i \eta}$, the Green's function is
\begin{equation} \label{eq:propeq}
G = \frac{1}{\omega - E_0 + i \eta} + \frac{1}{(\omega - E_0 + i \eta)^2} \bra{\psi}\tilde{T} \sum_{n=0}^{\infty}\left( \frac{\hat{P}}{\omega - \hat{H}_0 + i \eta} \tilde{T}\right)^n \ket{\psi}
\end{equation}
where 
\begin{equation}
\tilde{T} = \hat{V} \sum_{n=0}^{\infty} \left(\frac{\hat{Q}}{\omega - \hat{H}_0 + i \eta} \hat{V}\right)^n \ \ \ .
\end{equation}
\begin{equation}
\frac{\hat{P}}{\omega - \hat{H}_0 + i \eta} = \frac{\ket{\psi} \bra{\psi}}{\omega - E_0 + i \eta} \ \ \ ,
\end{equation}
so 
\begin{equation}
\sum_{n=0}^{\infty} \left( \frac{\hat{P}}{\omega - \hat{H}_0 + i \eta} \tilde{T} \right)^n \ket{\psi} = \sum_{n=0}^{\infty} \left( \frac{1}{\omega - E_0 + i \eta} \bra{\psi} \tilde{T} \ket{\psi} \right)^n \ket{\psi} \ \ \ .
\end{equation}
Using this fact, Eq. (\ref{eq:propeq}) becomes
$$
G = \frac{1}{\omega - E_0 + i \eta} + \frac{1}{(\omega - E_0 + i \eta)^2} \bra{\psi}\tilde{T}\ket{\psi} +  \frac{1}{(\omega - E_0 + i \eta)^3} \bra{\psi}\tilde{T}\ket{\psi}^2 + \cdots
$$
\begin{equation}
 = \frac{1}{\omega - E_0 + i \eta} \left(\frac{1}{1 - \frac{1}{\omega - E_0 + i \eta}\bra{\psi}\tilde{T}\ket{\psi}} \right) = \frac{1}{\omega - E_0 + i \eta - \bra{\psi}\tilde{T}\ket{\psi}} \ \ \ .
\end{equation}
In the limit that $\eta$ goes to zero, $\tilde{T}$ becomes the scattering $T$ matrix, $\hat{T}$, and according to the optical theorem $\frac{\Gamma}{2} = -\Im(\bra{\psi}\hat{T}\ket{\psi})$ for a single particle state, $\ket{\psi}$. So  
\begin{equation} \label{GreenResult}
\abs{\bra{\psi} \frac{1}{\omega - \hat{H} + i \eta} \ket{\psi}}^2 = \frac{1}{(\omega- E)^2 + (\frac{\Gamma}{2} + \eta)^2} \ \ \ ,
\end{equation}
and from Eq. (\ref{GreenResult}) the decay rate can be extracted since it is proportional to the width of a Lorentzian distribution centered at the particle's energy.

\section{Quantum Computation of the Green's Function}
\label{appendix:quantumgreen}
\subsection{Fully Quantum Approach}
\label{appendix:greencircuit}
In the previous section, it was shown that the imaginary part of the poles of the Green's function $\bra{\psi}\frac{1}{\omega - \hat{H} + i \eta}\ket{\psi} $ is $\frac{\Gamma}{2}  + \eta$. Therefore, if this Green's function can be computed efficiently, then the decay rate can be computed efficiently as well. This Green's function can be expressed as an integral
\begin{equation}
\bra{\psi} \frac{1}{\omega - \hat{H} + i \eta} \ket{\psi}
= - i\int_{0}^{\infty} e^{i (\omega + i \eta)t} \bra{\psi} e^{-i \hat{H} t} \ket{\psi} dt \ \ \ .
\end{equation}
If this integral is cut off  at some finite large time $T$, it can be approximated with a Riemann sum
\begin{equation}
 \lim_{T \rightarrow \infty, \Delta t \rightarrow 0}\sum_{k=0}^{T/\Delta t} e^{i(\omega + i \eta) k \Delta t} \bra{\psi} e^{-i \hat{H} k \Delta t} \ket{\psi} \Delta t = i \bra{\psi} \frac{1}{E - \hat{H} + i \eta} \ket{\psi} \ \ \ .
\end{equation}
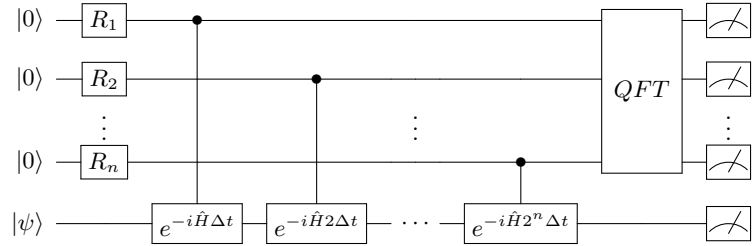
\begin{figure}[H]
\centering
\mbox{
\Qcircuit @C=1.em @R=1.em {
\lstick{\ket{0}} & \gate{R_1} & \ctrl{4} & \qw & \qw & \qw & \qw & \qw & \multigate{3}{QFT} & \meter \\
\lstick{\ket{0}} & \gate{R_2} & \qw  & \ctrl{3} & \qw & \qw & \qw & \qw & \ghost{QFT} & \meter \\
& \vdots  & & & &\vdots & & & & \vdots \\
\lstick{\ket{0}} & \gate{R_n}& \qw & \qw & \qw & \qw & \qw & \ctrl{1} & \ghost{QFT} &  \meter \\
\lstick{\ket{\psi}} & \qw & \gate{e^{-i \hat{H} \Delta t}} & \gate{e^{-i \hat{H} 2 \Delta t}} & \qw & \cdots &  & \gate{e^{-i \hat{H} 2^n \Delta t}} & \qw &  \meter
}
}

\caption{The quantum circuit used to calculate the Riemann sum approximation to the Green's function.}
\label{fig:circuit}
\end{figure}
If $T = (2^{n+1} - 1) \Delta t$, this sum can be evaluated on quantum computer using a register of n ancilla qubits in addition to a register used to store the state of the system and a number of gates that scale polynomially with $n$ and the size of the system. The circuit used to calculate the Green's function is displayed in Fig. \ref{fig:circuit}. The calculation begins with the quantum computer in the state $\ket{0}^{\otimes n} \ket{\psi}$ where all qubits in the ancilla are in the state $\ket{0}$ and the system register is in the state $\ket{\psi}$ which describes the unstable particle that will be decaying. $R_k$ is applied to the $k$th ancilla qubit, where
$
R_k = 
\begin{pmatrix} 
\cos(\theta_k) & -\sin(\theta_k) \\
\sin(\theta_k) & \cos(\theta_k) 
\end{pmatrix}
$
and $\theta_k = \arctan(e^{- 2^k \eta \Delta t})$. Up to normalization factors, the quantum computer is in the state $\sum \limits_{k=0}^{2^n-1} e^{-\eta k \Delta t}\ket{k} \ket{\psi}$. From the $k$th ancilla qubit, a controlled time evolution evolution operator is applied to the system register for time $2^k \Delta t $. Finally, the quantum Fourier transform is applied to the ancilla qubits which will put the quantum computer in the state $\sum \limits_{m=0}^{2^n - 1} \sum \limits_{k=0}^{2^n - 1} e^{i(\frac{2 \pi}{(2^n -1)\Delta t}m + i\eta) k \Delta t} \ket{m} e^{-i \hat{H} k \Delta t} \ket{\psi}$. Performing a measurement on both registers, the probability that the ancilla register is in the state $m$ and the system register is in the state $\psi$ is
\begin{equation}
P(m, \psi) \propto \abs{\sum_{k=0}^{2^n - 1} e^{i(\omega_m + i\eta) k \Delta t} \bra{\psi} e^{-i \hat{H} k \Delta t} \ket{\psi}}^2
\end{equation}
where $\omega_m = \frac{2 \pi m}{(2^{n+1} - 1) \Delta t}$. This is directly proportional to the Riemann sum that approximates the Green's function, and by repeatedly running this circuit, estimates for $P(m, \psi)$ can be obtained.

\subsection{Hybrid Approach}
\label{appendix:ampest}
\begin{figure}[H]
\centering
\subfloat[Circuit used to determine $ \Re(\bra{\psi} e^{-i H t} \ket{\psi})$]{
\mbox{
\Qcircuit @C=1em @R=1.em {
\lstick{\ket{0}}& \gate{H} & \ctrl{1} & \gate{H} & \meter \\
\lstick{\ket{\psi}}& \qw & \gate{e^{-i \hat{H} t}} & \qw & \meter
}
}
\label{fig:re}
}\hspace{0.05\textwidth} %
\subfloat[Circuit used to determine $ \Im(\bra{\psi} e^{-i H t} \ket{\psi})$]{
\mbox{
\Qcircuit @C=1em @R=1.em {
\lstick{\ket{0}}& \gate{H} & \ctrl{1} & \gate{S} & \gate{H} & \meter \\
\lstick{\ket{\psi}}& \qw & \gate{e^{-i \hat{H} t}} & \qw & \qw & \meter
}
}
\label{fig:im}
}

\caption{Circuits used in the Hadamard Test}

\end{figure}
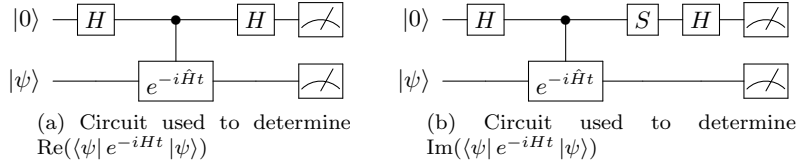

The circuit described in the previous section allows the Green's function to be computed using only quantum resources. However, that circuit requires many CNOT gates and ancilla qubits which makes implementation on a near term quantum computer difficult. Previous work has introduced variational methods to compute the Green's function on near term quantum computers \cite{GreenFnQC}. In this section, a method of computing Green's functions using the Hadamard test will be introduced. This method only requires a single ancilla qubit, which makes it more suitable for near term quantum computers than the method in the previous section. The Hadamard test method \cite{HadTest} can be used to compute $\bra{\psi} e^{-i \hat{H} t} \ket{\psi}$ with a quantum computer for several time slices. For the circuit in Fig. \ref{fig:re}, $P(0) - P(1) = \Re\left( \bra{\psi} e^{-i \hat{H} t} \ket{\psi}\right)$, where $P(0)$ is the probability that the ancilla qubit is measured to be in that state $0$ and $P(1)$ is the probability it is measured to be $1$. For the circuit in Fig. \ref{fig:im}, $P(0) - P(1) = \Im\left( \bra{\psi} e^{-i \hat{H} t} \ket{\psi}\right)$. By running these two circuits $n$ times, $\bra{\psi} e^{-i \hat{H} t} \ket{\psi}$ can be computed with statistical error given by $\frac{1}{\sqrt{n}}$.
Once $\bra{\psi} e^{-i \hat{H} t} \ket{\psi}$ has been computed for several time slices, the Green's function can be computed by classically performing a discrete Fourier transform. This requires a separate quantum circuit for each time slice, but the circuits used are shorter than the circuit in the previous section which makes them better suited for implementation on near term quantum computers. Implementing the Hadamard test requires at most a polynomial overhead over the cost of implementing $e^{-i H t}$. Therefore, the quantum and classical resources needed to compute the Green's function scale polynomially with this method as long as $\ket{\psi}$ can be prepared using polynomially many resources and time evolution can be performed using polynomially many resources on the quantum computer.

\section{Error Scaling}
\label{appendix:errors}
\subsection{Finite T and $\Delta t$}
\label{appendix:Terrors}
This calculation is based on performing a Riemann sum approximation to an integral, so errors due to a finite $T$ cutoff and a finite step size $\Delta t$ will need to be estimated. The Riemann sum being evaluated is
\begin{equation}
R = \sum_{k=0}^{T/\Delta t} e^{i(\omega + i \eta) k \Delta t} \bra{\psi} e^{-i \hat{H} k \Delta t} \ket{\psi} \Delta t
\end{equation}
which approximates 
\begin{equation} \label{eq:intcutoff}
I = \int_{0}^{T} e^{i (\omega + i \eta)t} \bra{\psi} e^{-i \hat{H} t} \ket{\psi} dt \ \ \ .
\end{equation} 
Eq. (\ref{eq:intcutoff}) differs from the $T \rightarrow \infty$ limit by
\begin{equation}
\delta I = \bra{\psi} \frac{1}{\omega - \hat{H} + i \eta} e^{i(\omega - \hat{H})T} \ket{\psi} e^{-\eta T} \ \ \ .
\end{equation}
Therefore, to determine the Green's function to within an accuracy of $\epsilon$, $T$ must be taken to be $O\left(\frac{1}{\eta} \log(\frac{1}{\eta \epsilon})\right)$. Using integration by parts, it can be shown that
\begin{equation}
\label{eq::ibperr}
\int_{t_1}^{t_2}dt e^{a t} = (t_2 - t_1) e^{a t_2} - \int _{t_1}^{t_2}dt \ (t - t_1) \ a \ e^{a t}
\end{equation}
and using Eq. \ref{eq::ibperr}, it can be shown that
\begin{equation}
I - R = -\sum_{n=1}^{n_t} \int_{\Delta t \ (n-1)}^{\Delta t \ n} dt \ (t - \Delta t \ (n-1)) \ \bra{\psi}(\omega - \hat{H} + i \eta) e^{i(\omega - \hat{H} + i \eta)t} \ket{\psi}
\end{equation}
where $n_t$ is the number of time slices used in the Riemann sum. $\ket{\psi}$ can be expanded in the eigenbasis of $\hat{H}$ as $\ket{\psi} = \sum_n c_n \ket{E_n}$. The Hamiltonians for which this method of computing the Green's function is to be applied to have been renormalized such that the lowest energy state has zero energy. Therefore, it may be assumed that $E_n \geq 0$ for all $n$. If $\bra{\psi} \hat{H} \ket{\psi} = E$, then
\begin{equation}
\abs{\bra{\psi} \hat{H} e^{-i \hat{H} t} \ket{\psi}} = \abs{\sum_n \abs{c_n}^2 E_n e^{-i E_n t}} \leq \sum_n \abs{c_n}^2 E_n = E \ \ \ .
\end{equation}
Using this bound, it can be shown that
\begin{equation}
\abs{I - R} \leq \int_{0}^{T} \Delta t \ (\omega + E + \eta) e^{-\eta t} \leq \frac{T}{\eta n_t} (\omega + \eta + E) \ \ \ .
\end{equation}
Therefore the number of time slices needed to determine the Green's function evaluated at $\omega + i \eta$ with an accuracy of $\epsilon$ must be
\begin{equation}
\text{Number of Time Slices} = O\left(\frac{\log(\frac{2}{\epsilon \eta})}{\eta^2 \epsilon}  \ (\omega + E + \eta)\right) \ \ \ .
\end{equation} 
Many implementations of Hamiltonian simulation on quantum computers do not implement the time evolution operator exactly \cite{QSim1,QSim2,QSim3,QSim4}. To calculate the Green's function with accuracy $\epsilon$, the error in the implementation of the time evolution operator must be $O(\eta \epsilon)$. If the gate cost required to evolve to a time $T$ with accuracy $\delta$ is given by $p(T,\delta)$, then the gate cost required to calculate the Green's function is 
\begin{equation}
\text{Gate Count} = O\left(\frac{\log(\frac{2}{\epsilon \eta})}{\eta^2 \epsilon} \ (\omega + E + \eta) \ p\left(\frac{1}{\eta} \log(\frac{2}{\eta \epsilon}),\eta \epsilon \right) \right) \ \ \ .
\end{equation}

\subsection{Finite Volume Errors}
\label{appendix:FVErr}
\subsubsection{$d+1$ Dimensions With a Mass Gap}
\label{appendix:FVErrMass}
When scattering calculations are done inside of a finite volume, first the $L \rightarrow \infty$ limit should be taken,  followed by the $\eta \rightarrow 0$ limit. The order in which this limit is taken matters, as can be seen from a rearrangement of $\Gamma = -2 \Im(\bra{\phi} \hat{T} \ket{\phi}) $,
\begin{equation}
\Gamma = \sum_n \frac{2 \eta}{(M-E_n)^2 + \eta^2} \abs{\bra{E_n} \hat{V} \ket{\phi} }^2
\end{equation}
where $\ket{E_n}$ are eigenstates of the full Hamiltonian, $\ket{\phi}$ is the state describing the unstable particle, $\hat{V}$ is the interaction piece of the Hamiltonian, and $M$ is the mass of the particle decaying. If the $\eta \rightarrow 0$ limit is taken first then this discrete sum goes to zero. Alternatively, if $L \rightarrow \infty$ first, the energy levels become continuous and 
\begin{equation}
\Gamma_\eta = \int \frac{2 \eta}{(M-E)^2 + \eta^2} \abs{\bra{E} \hat{V} \ket{\phi} }^2 \rho(E) dE \ \ \ .
\end{equation}
If then $\eta \rightarrow 0$, the Lorentzian term becomes a delta function and the usual expression for the decay rate is recovered.
\begin{equation}
\Gamma = 2 \pi \abs{\bra{M} \hat{V} \ket{\phi} }^2 \rho(M) \ \ \ .
\end{equation}
Finite volume errors in a $1 \rightarrow N$ particle decay rate will be calculated in the case where all of the decay products are massive. If the interaction energy between the decay products can be ignored (as in the $L \rightarrow \infty $ limit), then for a $1 \rightarrow N$ decay, the calculated decay rate is
\begin{equation} \label{FVGamma}
\Gamma_{FV, \eta} =  \sum_{\vec{n}_1, \vec{n}_2, \cdots \vec{n}_N \in \mathbb{Z}^d} \frac{2 \eta}{\left(M- \sum\limits_{k=1}^{N} \sqrt{m_k^2 + \left(\frac{2 \pi \vec{n}_k}{L}\right)^2}\right)^2 + \eta^2} \abs{\bra{\vec{n}_1, \vec{n}_2, \cdots \vec{n}_N} \hat{V} \ket{\phi} }^2
\end{equation}
in the $a \rightarrow 0$ limit, where $M$ is the mass of the heavy particle decaying and $m_k$ is the mass of the $k$-th decay product. In this case,
\begin{equation}
\abs{\bra{\vec{n}_1, \vec{n}_2, \cdots \vec{n}_N} \hat{V} \ket{\phi} }^2 = \int \frac{d^d \vec{x}}{L^d} e^{i \frac{2 \pi}{L} \vec{x} \cdot \left(\sum\limits_{k=1}^{N} \vec{n}_k \right)} \frac{\abs{\mathcal{M}}^2}{2 M} \prod\limits_{k=1}^{N} \frac{1}{L^d 2 \sqrt{m_k^2 + \left(\frac{2 \pi \vec{n}_k}{L} \right)^2}} 
\end{equation}
where $\mathcal{M}$ is the scattering amplitude which is generically an analytic function of all the decay products momentum, and $\vec{x}$ is integrated over the region $ [\frac{-L}{2}, \frac{L}{2} ]^d $. Therefore, the decay rate computed inside a finite volume at finite $\eta$ is
\begin{equation}
\label{GammaFV}
\Gamma_{FV, \eta} =  \sum_{\vec{n}_1, \vec{n}_2, \cdots \vec{n}_N \in \mathbb{Z}^d} \int \frac{d^d \vec{x}}{L^d} e^{i \frac{2 \pi}{L} \vec{x} \cdot \left(\sum\limits_{k=1}^{N} \vec{n}_k \right)} \frac{2 \eta}{\left(M- \sum\limits_{k=1}^{N} \sqrt{m_k^2 + \left(\frac{2 \pi \vec{n}_k}{L}\right)^2}\right)^2 + \eta^2} \frac{\abs{\mathcal{M}}^2}{2 M} \prod\limits_{k=1}^{N} \frac{1}{L^d 2 \sqrt{m_k^2 + \left(\frac{2 \pi \vec{n}_k}{L}\right)^2}}
\end{equation}
If instead, the goal is to calculate a cross section for $2 \rightarrow N$ scattering in the center of mass frame, an initial state with two particles each with energy $K_i$ must be prepared. In this case,
\begin{equation}
\abs{\bra{\vec{n}_1, \vec{n}_2, \cdots \vec{n}_N} V \ket{\phi} }^2 = \int \frac{d^d \vec{x}}{L^d} e^{i \frac{2 \pi}{L} \vec{x} \cdot \left(\sum\limits_{k=1}^{N} \vec{n}_k\right)} \frac{\abs{\mathcal{M}}^2}{4 K_i^2 L^d} \prod\limits_{k=1}^{N} \frac{1}{L^d 2 \sqrt{m_k^2 + \left(\frac{2 \pi \vec{n}_k}{L}\right)^2}} \ \ \ .  
\end{equation} 
The scattering cross section, $\sigma$, is given by the decay rate divided by the incident flux which is equal to $\frac{\abs{\vec{v_1}-\vec{v_2}}}{L^d} $, where $\vec{v_1}$ and $\vec{v_2}$ are the velocities of the particles present in the initial state. The extracted value for the cross section at finite volume and $\eta$ is given by
\begin{equation}
\sigma_{FV, \eta} = \sum_{\vec{n}_1, \vec{n}_2, \cdots \vec{n}_N \in \mathbb{Z}^d} \int \frac{d^d \vec{x}}{L^d} e^{i \frac{2 \pi}{L} \vec{x} \cdot \left(\sum\limits_{k=1}^{N} \vec{n}_k \right)} \frac{2 \eta}{\left(2 K_i - \sum\limits_{k=1}^{N} \sqrt{m_k^2 + \left(\frac{2 \pi \vec{n}_k}{L}\right)^2}\right)^2 + \eta^2} \frac{1}{\abs{\vec{v_1}-\vec{v_2}}} \frac{\abs{\mathcal{M}}^2}{4 K_i^2} \prod\limits_{k=1}^{N} \frac{1}{L^d 2 \sqrt{m_k^2 + \left(\frac{2 \pi \vec{n}_k}{L}\right)^2}} \ \ \ .
\end{equation}
This expression takes the same form as Eq. \ref{GammaFV} just with $M$ replaced by $2K_i$ and with some slightly different prefactors, so the finite volume error analysis for cross sections can proceed in the same way as the decay rate analysis. To simplify the following discussion, the finite volume errors will be computed only for decay rates.

As $L \rightarrow \infty$, the computed decay rate becomes
\begin{equation}
\label{GammaEpsilon}
\Gamma_{\eta} =   \prod\limits_{k=1}^{N} \int \frac{d^d \vec{p}_k}{(2\pi)^d 2 \sqrt{m_k^2 + \vec{p}_k^2}} \int d^d \vec{x} e^{i \vec{x} \cdot \left(\sum\limits_{k=1}^{N} \vec{p}_k \right)} \frac{2 \eta}{\left(M- \sum\limits_{k=1}^{N} \sqrt{m_k^2 + \vec{p}_k^2}\right)^2 + \eta^2} \frac{\abs{\mathcal{M}}^2}{2 M}
\end{equation}
The Poisson resummation formula states that
\begin{equation} \label{Poisson}
\sum_{\vec{n} \in \mathbb{Z}^d} f(\vec{n}) = \int d^d \vec{x} \ f(\vec{x}) + \sum_{\vec{p} \neq \vec{0}} \int d^d \vec{x} \ e^{2 \pi i  \vec{p} \cdot \vec{x}} f(\vec{x})
\end{equation}
and using Eq. (\ref{Poisson}), the finite volume error in the calculation of $\Gamma_\eta$ is given by
\begin{equation}
\delta \Gamma_{FV} = \Gamma_{FV,\eta} - \Gamma_{\eta} = -2\Im\left(\sum_{n \in \{ \{\vec{n_1,}, \vec{n_2}, \cdots, \vec{n_N} \in \mathbb{Z}^d \} \} \backslash \{\vec{0}, \cdots, \vec{0} \} } I_n \right)
\end{equation}
where
\begin{equation}
\label{ErrorIntegral1}
I_n = \int \frac{d^d \vec{x}}{L^d} \prod\limits_{k=1}^{N} \int \frac{d^d \vec{p}_k}{(2\pi)^d 2 \sqrt{m_k^2 + \vec{p}_k^2}} e^{i \vec{p}_k \cdot \left(\vec{n}_k L + \vec{x} \right)} \frac{1}{M- \sum\limits_{k=1}^{N} \sqrt{m_k^2 + \vec{p}_k^2} + i \eta} \frac{\abs{\mathcal{M}}^2}{2 M}  \ \ \ .
\end{equation}
Using the fact that 
\begin{equation}
\frac{1}{2 \sqrt{m^2 + \vec{p}^2}} = \int_\gamma \frac{dE}{2 \pi i} \frac{1}{(E + i \delta)^2 - \vec{p}^2 - m^2}
\end{equation}
where $\gamma$ is a contour enclosing the lower right quadrant of the complex plane, Eq. \ref{ErrorIntegral1} can be rewritten as
\begin{equation}
\label{ErrorIntegral2}
I_n = \int \frac{d^d \vec{x}}{L^d} \prod\limits_{k=1}^{N} \int_\gamma \frac{dE_k}{2 \pi i} \int \frac{d^d \vec{p}_k}{(2\pi)^d} \frac{e^{i \vec{p}_k \cdot \left(\vec{n}_k L + \vec{x} \right)}}{\left(E_k + i \frac{\eta}{N+1} \right)^2 - \vec{p}_k^2 - m_k^2} \frac{1}{M- \sum\limits_{k=1}^{N} E_k + i \frac{\eta}{N+1}} \frac{\abs{\mathcal{M}}^2}{2 M}  \ \ \ .
\end{equation}
The component of $\vec{p}_k$ parallel to $\vec{n}_k$ can be integrated over with contour integration yielding
\begin{equation}
\label{ErrorIntegral3}
I_n = \int \frac{d^d \vec{x}}{L^d} \prod\limits_{k=1}^{N} \int_\gamma \frac{dE_k}{2 \pi} \int \frac{d^{d-1} \vec{p}^T_k}{(2\pi)^{d-1}} \frac{e^{i \sqrt{(E_k + i \frac{\eta}{N+1})^2 - m_k^2 - (\vec{p}^T_k)^2} \abs{n_k L + \hat{n}_k \cdot \vec{x}} + i \vec{p}^T_k \cdot \vec{x}}}{2 \sqrt{(E_k + i \frac{\eta}{N+1})^2 - m_k^2 - (\vec{p}^T_k)^2}} \frac{1}{M- \sum\limits_{k=1}^{N} E_k + i \frac{\eta}{N+1}} \frac{\abs{\mathcal{M}}^2}{2 M}  
\end{equation}
where $p^T_k$ is a $d$-dimensional vector integrated over vectors perpendicular to $\vec{n}_k$. The integral over $p^T_k$ can be performed in the large $L$ limit using the saddle point approximation method which states that
\begin{equation}
\label{saddle}
\int d^d \vec{x} h(\vec{x}) e^{-\lambda f(\vec{x})} = \left(\frac{2 \pi}{\lambda} \right)^{\frac{d}{2}} h(\vec{x}_0) e^{-\lambda f(\vec{x}_0)} \frac{1}{\det(\text{Hessian}(f(\vec{x}_0)))^{\frac{1}{2}}}
\end{equation}
in the $\lambda \rightarrow \infty$ limit, where $\vec{x}_0$ is a stationary point of $f(\vec{x})$ in the integration domain. Therefore, in the large $L$ limit Eq. \ref{ErrorIntegral3} becomes
\begin{equation}
\label{ErrorIntegral4}
 I_n = \int \frac{d^d \vec{x}}{L^d} \prod\limits_{k=1}^{N} \int_\gamma \frac{dE_k}{4 \pi} \frac{1}{(2\pi)^{\frac{d-1}{2}}} \frac{\left(\sqrt{(E_k + i \frac{\eta}{N+1})^2 - m_k^2}\right)^{\frac{d-3}{2}}}{\abs{n_k L + \hat{n}_k \cdot \vec{x}}^{\frac{d-1}{2}}} e^{i \sqrt{(E_k + i \frac{\eta}{N+1})^2 - m_k^2} \abs{n_k L + \hat{n}_k \cdot \vec{x}}}  \frac{1}{M- \sum\limits_{k=1}^{N} E_k + i \frac{\eta}{N+1}} \frac{\abs{\mathcal{M}}^2}{2 M} \ \ \ .
\end{equation}
The $E_k$ integrals over $\gamma$ can be written as a sum over an integral over the positive real axis and the negative imaginary axis. Explicitly writing out these integrals,
\begin{multline}
I_n = \sum_{\sigma \subseteq \{1,2,\cdots,N\}} \int \frac{d^d \vec{x}}{L^d} \prod_{k \notin \sigma} \int_0^{\infty} \frac{dE_k}{4 \pi} \frac{1}{(2\pi)^{\frac{d-1}{2}}} \frac{\left(\sqrt{(E_k + i \frac{\eta}{N+1})^2 - m_k^2}\right)^{\frac{d-3}{2}}}{\abs{n_k L + \hat{n}_k \cdot \vec{x}}^{\frac{d-1}{2}}} e^{i \sqrt{(E_k + i \frac{\eta}{N+1})^2 - m_k^2} \abs{n_k L + \hat{n}_k \cdot \vec{x}}} \\
\prod_{k \in \sigma} \int_0^{\infty} \frac{-i dE_k}{4 \pi} \frac{1}{(2\pi)^{\frac{d-1}{2}}} \frac{\left(\sqrt{(E_k - \frac{\eta}{N+1})^2 + m_k^2}\right)^{\frac{d-3}{2}}}{\abs{n_k L + \hat{n}_k \cdot \vec{x}}^{\frac{d-1}{2}}} e^{- \sqrt{(E_k - \frac{\eta}{N+1})^2 + m_k^2} \abs{n_k L + \hat{n}_k \cdot \vec{x}}} \\
\frac{1}{M- \sum\limits_{k \notin \sigma} E_k + i \left(\frac{\eta}{N+1} +  \sum\limits_{k \in \sigma} E_k \right)} \frac{\abs{\mathcal{M}}^2}{2 M}
\end{multline}
For $k \in \sigma$, the $E_k$ integrals can be evaluated using the saddle point approximation again,
\begin{multline}
I_n = \sum_{\sigma \subseteq \{1,2,\cdots,N\}} \int \frac{d^d \vec{x}}{L^d} \prod_{k \notin \sigma} \int_0^{\infty} \frac{dE_k}{4 \pi} \frac{1}{(2\pi)^{\frac{d-1}{2}}} \frac{\left(\sqrt{(E_k + i \frac{\eta}{N+1})^2 - m_k^2}\right)^{\frac{d-3}{2}}}{\abs{n_k L + \hat{n}_k \cdot \vec{x}}^{\frac{d-1}{2}}} e^{i \sqrt{(E_k + i \frac{\eta}{N+1})^2 - m_k^2} \abs{n_k L + \hat{n}_k \cdot \vec{x}}} \\
\prod_{k \in \sigma} \frac{-i}{4 \pi} \frac{1}{(2\pi)^{\frac{d}{2}}} \frac{m_k^{\frac{d-2}{2}}}{\abs{n_k L + \hat{n}_k \cdot \vec{x}}^{\frac{d}{2}}} e^{- m_k \abs{n_k L + \hat{n}_k \cdot \vec{x}}} \\
\frac{1}{M- \sum\limits_{k \notin \sigma} E_k + i \eta \frac{\abs{\sigma} + 1}{N+1}} \frac{\abs{\mathcal{M}}^2}{2 M} \ \ \ .
\end{multline}
The final set of $E_k$ integrals will be performed by making the substitution $E_k = E f_k$,
\begin{multline}
I_n = \sum_{\sigma \subseteq \{1,2,\cdots,N\}} \int \frac{d^d \vec{x}}{L^d}  \int_0^{\infty} dE E^{N - \abs{\sigma}-1} \prod_{k \notin \sigma} \int \frac{df_k}{4 \pi} \frac{1}{(2\pi)^{\frac{d-1}{2}}} \frac{\left(\sqrt{(E f_k + i \frac{\eta}{N+1})^2 - m_k^2}\right)^{\frac{d-3}{2}}}{\abs{n_k L + \hat{n}_k \cdot \vec{x}}^{\frac{d-1}{2}}} e^{i \sqrt{(E f_k + i \frac{\eta}{N+1})^2 - m_k^2} \abs{n_k L + \hat{n}_k \cdot \vec{x}}} \\
\prod_{k \in \sigma} \frac{-i}{4 \pi} \frac{1}{(2\pi)^{\frac{d}{2}}} \frac{m_k^{\frac{d-2}{2}}}{\abs{n_k L + \hat{n}_k \cdot \vec{x}}^{\frac{d}{2}}} e^{- m_k \abs{n_k L + \hat{n}_k \cdot \vec{x}}} \frac{1}{M- E + i \eta \frac{\abs{\sigma} + 1}{N+1}} \frac{\abs{\mathcal{M}}^2}{2 M} 
\end{multline}
where the $f_k$ are integrated over the region $0 \leq f_k \leq 1$ and $\sum f_k = 1$. The integral over $E$ can be performed by performing a contour integration over a contour enclosing the upper quadrant of the complex plane. Performing this contour integral yields
\begin{equation}
I_n = \sum_{\sigma \subseteq \{1,2,\cdots,N\}} A_n(\sigma) + B_n(\sigma)
\end{equation}
where
\begin{multline}
\label{Aeq}
A_n(\sigma) = i \int \frac{d^d \vec{x}}{L^d} \prod_{k \notin \sigma} \int \frac{dE_k}{4 \pi} \frac{1}{(2\pi)^{\frac{d-1}{2}}} \frac{\left(\sqrt{(E_k +  \frac{\eta}{N+1})^2 + m_k^2}\right)^{\frac{d-3}{2}}}{\abs{n_k L + \hat{n}_k \cdot \vec{x}}^{\frac{d-1}{2}}} e^{- \sqrt{(E_k +  \frac{\eta}{N+1})^2 + m_k^2} \abs{n_k L + \hat{n}_k \cdot \vec{x}}} \\
\prod_{k \in \sigma} \frac{-i}{4 \pi} \frac{1}{(2\pi)^{\frac{d}{2}}} \frac{m_k^{\frac{d-2}{2}}}{\abs{n_k L + \hat{n}_k \cdot \vec{x}}^{\frac{d}{2}}} e^{- m_k \abs{n_k L + \hat{n}_k \cdot \vec{x}}} \frac{1}{M- \sum_{k \notin \sigma}i E_k + i \eta \frac{\abs{\sigma} + 1}{N+1}} \frac{\abs{\mathcal{M}}^2}{2 M}
\end{multline}
comes from integrating along the positive imaginary axis and
\begin{multline}
B_n(\sigma) =  -2 \pi i \left(M + i \eta \frac{\abs{\sigma} + 1}{N+1}\right)^{N - \abs{\sigma}-1} \prod_{k \notin \sigma} \int \frac{df_k}{4 \pi} \frac{1}{(2\pi)^{\frac{d-1}{2}}} \frac{\left(\sqrt{\left(\left(M + i \eta \frac{\abs{\sigma} + 1}{N+1}\right) f_k + i \frac{\eta}{N+1}\right)^2 - m_k^2}\right)^{\frac{d-3}{2}}}{\abs{n_k L + \hat{n}_k \cdot \vec{x}}^{\frac{d-1}{2}}} \\ e^{i \sqrt{\left(\left(M + i \eta \frac{\abs{\sigma} + 1}{N+1}\right) f_k + i \frac{\eta}{N+1}\right)^2 - m_k^2} \abs{n_k L + \hat{n}_k \cdot \vec{x}}} \\
\prod_{k \in \sigma} \frac{-i}{4 \pi} \frac{1}{(2\pi)^{\frac{d}{2}}} \frac{m_k^{\frac{d-2}{2}}}{\abs{n_k L + \hat{n}_k \cdot \vec{x}}^{\frac{d}{2}}} e^{- m_k \abs{n_k L + \hat{n}_k \cdot \vec{x}}} \frac{\abs{\mathcal{M}}^2}{2 M}
\end{multline}
comes from the pole located at $E = M + i \eta \frac{\abs{\sigma} + 1}{N+1}$
The integrals in Eq. \ref{Aeq} can be evaluated using the saddle point approximation and using the fact that $\abs{n_k L + \hat{n}_k \cdot \vec{x}} \geq (n_k - \frac{1}{2})L$, it follows that $A_n(\sigma) = O(\frac{1}{M^2}\prod\limits_k e^{-m_k (n_k - \frac{1}{2})L})$. Using the bound
\begin{equation}
\Im\left(\sqrt{\left(\left(M + i \eta \frac{\abs{\sigma} + 1}{N+1}\right) f_k + i \frac{\eta}{N+1}\right)^2 - m_k^2}\right) \geq  \frac{\eta}{N+1} \ \ \ ,
\end{equation}
it can be seen that
\begin{equation}
B_n(\sigma) = O\left(M^{\frac{d-1}{2}N-2}\prod\limits_{k \in \sigma} e^{-m_k (n_k - \frac{1}{2})L} \prod\limits_{k \notin \sigma} e^{-\frac{\eta}{N+1} (n_k - \frac{1}{2})L}\right) \ \ \ .
\end{equation}
Therefore, in the limit of small $\eta$, 
\begin{equation}
I_n = O\left(M^{\frac{d-1}{2}N-2} \prod\limits_{k=1}^{N} e^{-\frac{\eta}{N+1} (n_k - \frac{1}{2})L}\right) \ \ \ ,
\end{equation}
and
\begin{equation}
\delta \Gamma_{FV} = O\left(M^{\frac{d-1}{2}N-2} e^{-\frac{\eta}{N+1}\frac{L}{2}}\right) \ \ \ .
\end{equation}

\subsubsection{$d+1$ Dimensions Without a Mass Gap}
\label{appendix:generalFV}
The bound in the previous section was calculated under the assumption that all decay products are massive, however this is not always the case. In the case where there are massless particles, the decay rate calculated in finite volume is again given by Eq. \ref{FVGamma}, and the infinite volume limit is given by Eq. \ref{GammaEpsilon}. $\Gamma_{FV,\eta}$ is a Riemann sum approximation to $\Gamma_\eta$ and using the multidimensional bound on the Riemann sum error for integrating over a $k$ dimensional hypercube with side length $R$
\begin{equation}
\abs{\int d^k x \ f(\vec{x}) - \sum_{\vec{n} \in \mathbb{Z}^n \backslash \{\abs{n_i} \Delta x \geq \frac{R}{2} \} } \Delta x^n  \ f(\Delta x \vec{n})} \leq \ \frac{\sum_{i=1}^{k} \max\abs{\frac{\partial f}{\partial x_i}}}{2} \  R^k \ \Delta x \ \ \ ,
\end{equation}
it can be seen that $\delta \Gamma_{FV}$ is $O(\frac{1}{\eta^2 L})$ for small $\eta$.

\subsection{Finite $\eta$  Errors}
\label{appendix:epserror}
The value of $\Gamma$ calculated for a $1 \rightarrow N$ particle decay at finite $\eta$ in an infinite volume is given by ${\Gamma_{\eta} = -2\Im(T(M + i \eta))}$ where
\begin{equation} \label{forwardamp}
T(z) =  \prod\limits_{k=1}^{N} \int \frac{d^d \vec{p}_k}{(2\pi)^d 2 \sqrt{m_k^2 + \vec{p}_k^2}} \frac{(2\pi)^d \delta^d\left(\sum\limits_{k=1}^{N} \vec{p}_k\right)}{z - \sum\limits_{k=1}^{N} \sqrt{m_k^2 + \vec{p}_k^2}} \frac{\abs{\mathcal{M}}^2}{2 M}  
\end{equation}
is the forward scattering amplitude, $M$ is the mass of the particle decaying, $m_k$ is the mass of the k-th decay product and $\mathcal{M}$ is the scattering amplitude for the given decay channel. The decay rate $\Gamma$ is given by $\lim \limits_{\eta \rightarrow 0} \Gamma_\eta $. It will be shown in this section that $\delta \Gamma_\eta = \Gamma_\eta - \Gamma $ is $O(\eta)$ for small $\eta$. Changing to spherical coordinates and making the substitution $E_k = \sqrt{p_k^2 + m_k^2} - m_k$, Eq. \ref{forwardamp} becomes
\begin{equation}
T(z) =  \prod\limits_{k=1}^{N} \int_{0}^{\infty} dE_k \int d \Omega_k \frac{(E_k^2 + 2 m_k E_k)^{\frac{d-2}{2}}}{2 (2\pi)^d} \frac{(2\pi)^d \delta^d \left(\sum\limits_{k=1}^{N} \vec{p}_k \right)}{z - \sum\limits_{k=1}^{N} (E_k + m_k)} \frac{\abs{\mathcal{M}}^2}{2 M}  \ \ \ .
\end{equation}
Now making the substitution $E_k = f_k E$, $T(z)$ can be expressed in the form
\begin{equation}
T(z) = \int_{0}^{\infty} dE f(E) \frac{1}{z - E - \sum_{k=1}^{N} m_k}
\end{equation}
where
\begin{equation}
f(E) = E^{N-1} \prod\limits_{k=1}^{N} \int df_k \int d \Omega_k \frac{(E^2 f_k^2 + 2 m_k E f_k)^{\frac{d-2}{2}}}{2 (2\pi)^d} (2\pi)^d \delta^d \left(\sum\limits_{k=1}^{N} \vec{p}_k \right) \frac{\abs{\mathcal{M}}^2}{2 M}  \ \ \ .
\end{equation}
where the $f_k$ are integrated over the region $0 \leq f_k \leq 1$ and $\sum_{k=1}^{N} f_k = 1$. Note that $f(E)$ has the following properties, $2\pi f(\Delta M) = \Gamma $ where $\Delta M = M - \sum_{k=1}^{N} m_k$, $f(0) = 0$, and $f(E) \geq 0$. $T(z)$ is known to be analytic in the upper half of the complex plane \cite{Schwartz} which implies $ \lim\limits_{E \rightarrow \infty}f(E)=0$, since otherwise $\Re(T(z))$ would diverge. The decay rate calculated at finite $\eta$ is
\begin{equation}
\label{LorentzianInt}
\Gamma_{\eta} = \int_{0}^{\infty} dE f(E) \frac{2 \eta}{(\Delta M - E)^2 + \eta^2} \ \ \ .
\end{equation}
Integrating Eq. \ref{LorentzianInt} by parts gives
\begin{equation}
\Gamma_{\eta} = -2 \int_{0}^{\infty} dE f'(E) \tan^{-1}\left(\frac{E - \Delta M}{\eta}\right) = 2\int_{\infty}^{\Delta M} dE f'(E) \cot^{-1}\left(\frac{\eta}{E - \Delta M}\right) - 2\int_{0}^{\Delta M} dE f'(E) \cot^{-1}\left(\frac{\eta}{E - \Delta M}\right)
\end{equation}
To show that $\delta \Gamma_\eta = \Gamma_\eta - \Gamma$ is $O(\eta)$ for small $\eta$, it suffices to show that
\begin{equation}
\lim\limits_{\eta \rightarrow 0^+}\left(\frac{\delta \Gamma_\eta}{\eta}\right) = \frac{d \Gamma_\eta}{d \eta}\Bigr|_{\substack{\eta = 0^+}}
\end{equation}
is finite. Differentiating under the integral shows that for $\eta > 0$,
\begin{equation}
\frac{d \Gamma_\eta}{d \eta} = 2 \int_{0}^{\infty} dE f'(E) \frac{E - \Delta M}{\eta^2 + (E - \Delta M)^2} = 2 \mathcal{P} \int_{0}^{\infty} dE f'(E) \frac{E - \Delta M}{\eta^2 + (E - \Delta M)^2} \ \ \ .
\end{equation}
Therefore,
\begin{equation}
\label{dGamEps}
\frac{d \Gamma_\eta}{d \eta}\Bigr|_{\substack{\eta = 0^+}} = 2 \mathcal{P} \int_{0}^{\infty} dE f'(E) \frac{1}{E - \Delta M} \ \ \ .
\end{equation}
The integral in Eq. \ref{dGamEps} is finite due to the properties of $f(E)$ discussed above and so $\delta \Gamma_\eta$ is $O(\eta)$ for small $\eta$.

\subsection{Particle Number Truncation}
\label{appendix:ParticleTruncation}
When a quantum field theory describing bosons is simulated on a quantum computer, the degrees of freedom must be truncated. For the bosonic theories considered in this paper, this was done by simulating the theory on a finite lattice with particle numbers truncated. The calculations in the previous section bounded the error in the computed decay rate due to the finite lattice and in this section, the error due to the particle number truncation will be calculated. The scattering $T$ matrix can be computed from the recurrence relation
\begin{equation}
\hat{T} = \hat{V} + \hat{V}\frac{1}{E - \hat{H}_0 + i \eta} \hat{T}
\end{equation}
where $V_0$ is the free part of the Hamiltonian describing the motion of free particles and $\hat{V}$ is the interaction part of the Hamiltonian. If $\hat{P}$ projects out the finite particle subspace under consideration, then the $T$ matrix computed with this truncation satisfies the recurrence relation
\begin{equation}
\hat{T}_f = \hat{V} + \hat{V} \frac{\hat{P}}{E - \hat{H}_0 + i \eta} \hat{T}_f
\end{equation}
Then the difference between the actual $T$ matrix and the $T$ matrix computed with a particle number truncation, $\hat{\delta} = \hat{T} - \hat{T}_f$ satisfies
\begin{equation}
\hat{\delta} = \hat{V}\frac{1}{E-\hat{H}_0 + i \eta} \hat{\delta} + \hat{V}\frac{1 - \hat{P}}{E - \hat{H}_0 + i \eta} \hat{T}_f \ \ \ .
\end{equation}
This can be rewritten as
\begin{equation}
\hat{\delta} = \hat{T} \frac{1 - \hat{P}}{E - \hat{H}_0 + i \eta} \hat{T}_f \ \ \ .
\end{equation}
Therefore, if the lightest particle in the theory has mass $m$ and particle number is truncated at $n$, then the error in $T(E)$ due to the particle number truncation is $O\left(\frac{1}{E - m(n+1)}\right)$. 
\section{Error Mitigation}
\label{appendix:mitigation}
While the Hadamard test enables the computation of matrix elements, it does not address errors due to imperfect gates on the device itself. To mitigate this error, an extrapolation technique was used \cite{errorextrap1,errorextrap2}. In each circuit, every CNOT was replaced with an odd number, $r$ (for $r=3,5,7$), of CNOT's, and each amplitude was linearly extrapolated to $r=0$. If there was no noise, these additional CNOT gates would make no change to the outcome of the circuit.

This procedure reduces the error from imperfect implementation of CNOT gates on the quantum computer, but does not mitigate readout errors. To address readout errors, the default calibration matrix method included in the Qiskit Ignis package was used \cite{Qiskit}.

\section{Hamiltonian Simulation}
\label{appendix:hamsim} 
The one site calculation done on IBM's quantum computer was done in the momentum basis. While the gate cost of performing time evolution in the momentum basis does not scale to large lattices as well as in the position basis, it is suitable for small calculations \cite{ScalarBenchmark}. With a single site,

$$\hat{\phi} = \frac{1}{\sqrt{2 M }}(\hat{a}_\phi + \hat{a}_\phi^\dagger) $$
$$\hat{\chi} = \frac{1}{\sqrt{2 m }}(\hat{a}_\chi + \hat{a}_\chi^\dagger)$$
$$\hat{\pi}_\phi = - i \sqrt{\frac{M}{2}} (\hat{a}_\phi - \hat{a}_\phi^\dagger) $$
$$\hat{\pi}_\chi = - i \sqrt{\frac{m}{2}} (\hat{a}_\chi - \hat{a}_\chi^\dagger)$$
\begin{equation}
\hat{H} = \frac{1}{2}\hat{\pi}_\phi^2 + \frac{1}{2}\hat{\pi}_\chi^2 + \frac{1}{2}M^2 \hat{\phi}^2 + \frac{1}{2} m^2 \hat{\chi}^2 + \frac{1}{2} g \ \hat{\phi} \ \hat{\chi}^2 + \frac{1}{4!} \lambda \ \hat{\chi}^4 + \frac{1}{2} \delta M^2 \hat{\phi}^2 + \frac{1}{2} \delta m^2 \hat{\chi}^2 + \Lambda \ \ \ .
\end{equation}
where $H$ is the Hamiltonian, $M$ is the mass of the heavy particle, $m$ is the mass of the light particle, $\Lambda$ is chosen to make the vacuum energy equal to zero, and $\delta M$ and $\delta m$ are the differences between the physical and bare masses. This Hamiltonian only couples states with the same parity in the number of $\chi$ particles so states with an even number of $\chi$ particles are the only ones needed. The mapping of  basis states to qubit states is listed in Table \ref{tab:basis}. Two qubits were used to store the state of the system and one ancilla qubit was used to implement the amplitude estimation algorithm described in Appendix \ref{appendix:ampest}.

\begin{table}[h]
\centering
\begin{tabular}{|l|l|}
\hline
Qubit State & Basis State\\
\hline
00 & Vacuum \\
01 & 1 $\phi$ \\
10 & 2 $\chi$ \\
11 & 1 $\phi$ and 2 $\chi$ \\
\hline
\end{tabular}
\caption{Basis States}
\label{tab:basis}
\end{table}

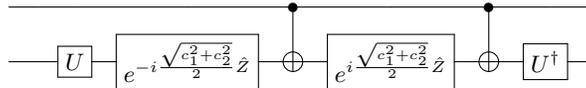
\begin{figure}[H]
\centering
\mbox{
\Qcircuit @C=1em @R=1.em {
& \qw & \qw & \qw & \ctrl{1} & \qw & \ctrl{1} & \qw & \qw \\
& \qw & \gate{U} & \gate{e^{-i \frac{\sqrt{c_1^2 + c_2^2}}{2} \hat{Z}}} & \targ & \gate{e^{i \frac{\sqrt{c_1^2 + c_2^2}}{2} \hat{Z}}} & \targ & \gate{U^\dagger} & \qw
}}
\caption{Circuit for $e^{-i(c_1 \hat{X} + c_2 \hat{Z})}$ controlled on the first qubit}
\label{fig:XZ}

\end{figure}

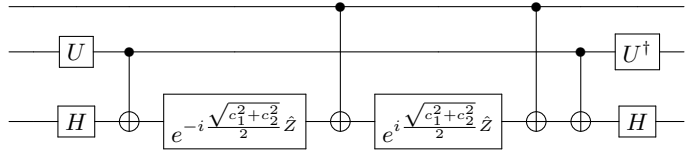
\begin{figure}[H]
\centering
\mbox{
\Qcircuit @C=1em @R=1.em {
& \qw & \qw & \qw & \qw & \ctrl{2} & \qw & \ctrl{2} & \qw & \qw & \qw \\
& \qw & \gate{U} & \ctrl{1} & \qw & \qw & \qw & \qw  & \ctrl{1} & \gate{U^\dagger} & \qw \\
& \qw & \gate{H} & \targ & \gate{e^{-i \frac{\sqrt{c_1^2 + c_2^2}}{2}  \hat{Z}}} & \targ & \gate{e^{i \frac{\sqrt{c_1^2 + c_2^2}}{2} \hat{Z}}} & \targ & \targ & \gate{H} & \qw
}}
\caption{Circuit for $e^{-i(c_1 \hat{X} + c_2 \hat{Z}) \otimes \hat{X}}$ controlled on the first qubit}
\label{fig:XZX}

\end{figure}

In this truncated basis, the Hamiltonian is
\begin{multline}
\\
\bar{H} = \left(\frac{M}{2} + m + \frac{7 \lambda}{32 m^2 } + \frac{\delta M^2}{2 M} + \frac{3 \delta m^2}{4 m} +  \Lambda \right) \hat{1} \otimes \hat{1} \\
+ \left(\left(\frac{\lambda}{8\sqrt{2}m^2 } + \frac{\sqrt{2} \delta m^2}{4m}\right) \hat{X} - \left(m + \frac{3 \lambda}{16 m^2 } + \frac{\delta m^2}{2 m}\right) \hat{Z}\right)\otimes \hat{1} \\
 + \hat{1} \otimes \left(\frac{3 g}{4 m \sqrt{2 M }}\hat{X} - \left(\frac{M}{2} + \frac{\delta M^2}{4 M}\right) \hat{Z}\right)  \\
 + \left(\frac{g}{4 m \sqrt{M}} \hat{X} - \frac{g}{2m \sqrt{2 M }} \hat{Z}\right)\otimes \hat{X} \ \ \ . \\
\end{multline}
The amplitude estimation procedure described in the previous section requires implementation of a controlled time evolution operator which was implemented using a Trotter-Suzuki decomposition $e^{-i \sum H_k \delta t} \approx \prod_k e^{-i H_k \delta t}$ where 
$$\hat{H}_0 = \left(\frac{M}{2} + m + \frac{7 \lambda}{32 m^2 } + \frac{\delta M^2}{2 M} + \frac{3 \delta m^2}{4 m} +  \Lambda \right) \hat{1} \otimes \hat{1} $$
$$\hat{H}_1 =  \left(\left(\frac{\lambda}{8\sqrt{2}m^2 } + \frac{\sqrt{2} \delta m^2}{4m}\right) \hat{X} - (m + \frac{3 \lambda}{16 m^2 } + \frac{\delta m^2}{2 m}) \hat{Z}\right)\otimes \hat{1}$$
$$\hat{H}_2 = \hat{1} \otimes \left(\frac{3 g}{4 m \sqrt{2 M }}\hat{X} - \left(\frac{M}{2} + \frac{\delta M^2}{4 M} \right) \hat{Z}\right)$$
\begin{equation}
\hat{H}_3=\left(\frac{g}{4 m \sqrt{M }} \hat{X} - \frac{g}{2m \sqrt{2 M }} \hat{Z}\right)\otimes \hat{X} \ \ \ .
\end{equation}
Implementing a controlled version of this time evolution operator requires the ability to perform controlled unitary transformations of the form $e^{i c \hat{1} \otimes \hat{1}} $, $e^{i(c_1 \hat{X} + c_2 \hat{Z}) \otimes \hat{1}} $ and $e^{i(c_1 \hat{X} + c_2 \hat{Z}) \otimes \hat{X}}$. A controlled $e^{i c \hat{1} \otimes \hat{1}} $ can be performed by applying $ 
\begin{pmatrix} 
1 & 0 \\
0 & e^{i c} 
\end{pmatrix}
$
to the control qubit. $c_1 \hat{X} + c_2 \hat{Z}$ is a 2x2 Hermitian matrix, and the matrix $U$ that maps the computational basis to the eigenbasis of this matrix can be found classically. Using this matrix $U$, it is trivial to modify the textbook implementation of a controlled $\hat{Z}$ rotation \cite{Nielsen:2011:QCQ:1972505} to a controlled rotation about $c_1 \hat{X} + c_2 \hat{Z}$ as shown in Fig. \ref{fig:XZ}. A similar trick can be used to implement the $(c_1 \hat{X} + c_2 \hat{Z}) \otimes \hat{X}$ term as shown in Fig. \ref{fig:XZX}.

\section{Estimation of Imperfect Gate Implementation Errors}
\label{appendix:gateerr}
On NISQ era quantum computers, the statistical error and error due to imperfect implementation of logic gates on the quantum processor must both be addressed. In general, the density matrix describing the state of the quantum computer is given by
\begin{equation}
\rho_{exp} = (1-p) \rho_{ideal} + \sum_i E_i \rho_{ideal} E_i^\dagger
\end{equation}
where $\rho_{ideal}$ is the density matrix describing the state of the quantum computer if every gate was implemented perfectly, $p$ is the probability there is an error anywhere in the circuit, $E_i$ are the Krauss operators describing the errors and $\sum \limits_i E_i E_i^\dagger = p$. The difference between the probability observed on a real quantum computer and an ideal quantum computer is given by
\begin{equation}
Tr\left(-p \ O \ \rho_{ideal} + \sum_i O \ E_i \ \rho_{ideal} \ E_i^\dagger \right) = p \ Tr \left(\frac{O}{p} \ \sum_i E_i \ \rho_{ideal} \ E_i^\dagger - O \ \rho_{ideal} \right)
\end{equation}
where $O$ is the projection operator corresponding to the measurement result. $\frac{1}{p}\sum \limits_i E_i \rho_{ideal} E_i^\dagger$ is a density matrix because $\sum \limits_i E_i E_i^\dagger = p$. So $Tr \left(\frac{O}{p}\sum \limits_i E_i \ \rho_{ideal} \ E_i^\dagger - O \rho_{ideal} \right)$ is a difference of probabilities which must be bounded above by one. As a result, the difference between the probability of a given measurement observed on a real quantum computer and an ideal quantum computer is bounded above by $p$. For the calculation on IBM's Ourense quantum processor, $p$ was calculated using the calibration data provided by IBM, and was used as an estimate of the error due to imperfect gate implementation.

\section{Data}
\label{appendix:data}
The following tables contain the results of all computations run on the IBM Ourense quantum processor.
\begin{longtable}{| c | c | c | c | c |} 
\hline 
\multicolumn{5}{|c|}{Error Mitigated Real Hadamard Test P(0) g = 0.5} \\ 
\hline 
Time Slice & r=1 & r=3 & r=5 & r=7 \\ 
\hline 
1 & 0.694 & 0.341 & 0.195 & 0.152 \\  
2 & 0.608 & 0.279 & 0.163 & 0.137 \\  
3 & 0.492 & 0.208 & 0.152 & 0.132 \\  
4 & 0.342 & 0.161 & 0.139 & 0.115 \\  
5 & 0.194 & 0.123 & 0.109 & 0.114 \\  
6 & 0.091 & 0.094 & 0.105 & 0.111 \\  
7 & 0.057 & 0.108 & 0.119 & 0.111 \\  
8 & 0.053 & 0.108 & 0.121 & 0.113 \\  
9 & 0.116 & 0.144 & 0.142 & 0.121 \\  
10 & 0.21 & 0.195 & 0.149 & 0.114 \\  
11 & 0.337 & 0.264 & 0.183 & 0.125 \\  
12 & 0.475 & 0.336 & 0.204 & 0.133 \\  
13 & 0.55 & 0.345 & 0.233 & 0.137 \\  
14 & 0.641 & 0.357 & 0.236 & 0.159 \\  
15 & 0.683 & 0.354 & 0.241 & 0.15 \\  
16 & 0.674 & 0.371 & 0.265 & 0.185 \\  
17 & 0.633 & 0.312 & 0.233 & 0.16 \\  
18 & 0.552 & 0.266 & 0.205 & 0.122 \\  
19 & 0.446 & 0.236 & 0.141 & 0.12 \\  
20 & 0.37 & 0.207 & 0.11 & 0.117 \\  
21 & 0.278 & 0.147 & 0.101 & 0.107 \\  
22 & 0.196 & 0.107 & 0.095 & 0.112 \\  
23 & 0.141 & 0.087 & 0.111 & 0.113 \\  
24 & 0.107 & 0.082 & 0.107 & 0.129 \\  
25 & 0.084 & 0.112 & 0.097 & 0.111 \\  
26 & 0.108 & 0.127 & 0.096 & 0.123 \\  
27 & 0.144 & 0.145 & 0.118 & 0.112 \\  
28 & 0.189 & 0.174 & 0.124 & 0.113 \\  
29 & 0.234 & 0.177 & 0.141 & 0.137 \\  
30 & 0.289 & 0.172 & 0.155 & 0.132 \\  
31 & 0.272 & 0.195 & 0.16 & 0.13 \\  
32 & 0.218 & 0.162 & 0.13 & 0.117 \\  
33 & 0.169 & 0.126 & 0.134 & 0.12 \\  
34 & 0.131 & 0.108 & 0.13 & 0.129 \\  
35 & 0.087 & 0.114 & 0.123 & 0.124 \\  
36 & 0.072 & 0.117 & 0.131 & 0.135 \\  
37 & 0.116 & 0.148 & 0.172 & 0.139 \\  
38 & 0.206 & 0.195 & 0.19 & 0.155 \\  
39 & 0.311 & 0.251 & 0.206 & 0.161 \\  
40 & 0.379 & 0.254 & 0.172 & 0.143 \\  
41 & 0.447 & 0.262 & 0.176 & 0.143 \\  
42 & 0.524 & 0.276 & 0.186 & 0.144 \\  
43 & 0.555 & 0.284 & 0.186 & 0.15 \\  
44 & 0.564 & 0.256 & 0.164 & 0.132 \\  
45 & 0.512 & 0.245 & 0.162 & 0.132 \\  
46 & 0.433 & 0.2 & 0.157 & 0.131 \\  
47 & 0.345 & 0.199 & 0.139 & 0.118 \\  
48 & 0.243 & 0.165 & 0.139 & 0.114 \\  
49 & 0.19 & 0.135 & 0.123 & 0.112 \\  
50 & 0.14 & 0.121 & 0.126 & 0.111 \\  
51 & 0.142 & 0.125 & 0.134 & 0.112 \\  
52 & 0.17 & 0.126 & 0.13 & 0.123 \\  
53 & 0.214 & 0.134 & 0.13 & 0.116 \\  
54 & 0.248 & 0.147 & 0.121 & 0.117 \\  
55 & 0.288 & 0.127 & 0.126 & 0.119 \\  
56 & 0.267 & 0.124 & 0.1 & 0.112 \\  
57 & 0.213 & 0.108 & 0.101 & 0.11 \\  
58 & 0.167 & 0.104 & 0.113 & 0.106 \\  
59 & 0.121 & 0.094 & 0.113 & 0.117 \\  
60 & 0.084 & 0.101 & 0.117 & 0.129 \\  
61 & 0.046 & 0.089 & 0.112 & 0.141 \\  
62 & 0.052 & 0.117 & 0.129 & 0.145 \\  
63 & 0.075 & 0.174 & 0.155 & 0.151 \\  
64 & 0.136 & 0.222 & 0.173 & 0.141 \\  
65 & 0.238 & 0.267 & 0.197 & 0.151 \\  
66 & 0.313 & 0.282 & 0.208 & 0.139 \\  
67 & 0.423 & 0.251 & 0.202 & 0.136 \\  
68 & 0.473 & 0.291 & 0.208 & 0.156 \\  
69 & 0.507 & 0.291 & 0.235 & 0.15 \\  
70 & 0.543 & 0.297 & 0.211 & 0.131 \\  
71 & 0.579 & 0.283 & 0.188 & 0.12 \\  
72 & 0.571 & 0.291 & 0.178 & 0.126 \\  
73 & 0.525 & 0.306 & 0.198 & 0.146 \\  
74 & 0.453 & 0.315 & 0.204 & 0.153 \\  
75 & 0.385 & 0.273 & 0.212 & 0.145 \\  
76 & 0.29 & 0.232 & 0.195 & 0.157 \\  
77 & 0.208 & 0.173 & 0.165 & 0.133 \\  
78 & 0.133 & 0.131 & 0.136 & 0.121 \\  
79 & 0.072 & 0.09 & 0.115 & 0.107 \\  
80 & 0.045 & 0.089 & 0.105 & 0.103 \\  
81 & 0.072 & 0.094 & 0.109 & 0.105 \\  
82 & 0.197 & 0.186 & 0.209 & 0.146 \\  
83 & 0.35 & 0.251 & 0.238 & 0.14 \\  
84 & 0.496 & 0.321 & 0.251 & 0.147 \\  
85 & 0.635 & 0.394 & 0.257 & 0.156 \\  
86 & 0.689 & 0.441 & 0.258 & 0.159 \\  
87 & 0.698 & 0.425 & 0.246 & 0.155 \\  
88 & 0.645 & 0.4 & 0.246 & 0.17 \\  
89 & 0.558 & 0.305 & 0.237 & 0.161 \\  
90 & 0.469 & 0.238 & 0.199 & 0.15 \\  
91 & 0.343 & 0.187 & 0.151 & 0.122 \\  
92 & 0.202 & 0.132 & 0.124 & 0.126 \\  
93 & 0.109 & 0.108 & 0.107 & 0.11 \\  
94 & 0.051 & 0.093 & 0.125 & 0.121 \\  
95 & 0.062 & 0.103 & 0.14 & 0.124 \\  
\hline
\caption{The probability of measuring zero in the ancilla qubit for the Hadamard test to determine the real part of $\bra{\psi} e^{-i H t} \ket{\psi}$ for $g = 0.5 $ after applying the measurement noise mitigation procedure described in Appendix \ref{appendix:mitigation} for different numbers of CNOT gates. Each entry in this table was calculated with $8192$ measurements.}
\end{longtable} 
 
 \begin{longtable}{| c | c | c | c | c |} 
\hline 
\multicolumn{5}{|c|}{Error Mitigated Real Hadamard Test P(1) g = 0.5} \\ 
\hline 
Time Slice & r=1 & r=3 & r=5 & r=7 \\ 
\hline 
1 & 0.07 & 0.142 & 0.138 & 0.122 \\  
2 & 0.162 & 0.197 & 0.18 & 0.125 \\  
3 & 0.266 & 0.256 & 0.206 & 0.144 \\  
4 & 0.397 & 0.31 & 0.208 & 0.166 \\  
5 & 0.523 & 0.364 & 0.24 & 0.171 \\  
6 & 0.612 & 0.373 & 0.248 & 0.185 \\  
7 & 0.68 & 0.38 & 0.259 & 0.194 \\  
8 & 0.666 & 0.369 & 0.241 & 0.179 \\  
9 & 0.601 & 0.349 & 0.223 & 0.171 \\  
10 & 0.497 & 0.272 & 0.209 & 0.15 \\  
11 & 0.374 & 0.203 & 0.154 & 0.125 \\  
12 & 0.256 & 0.136 & 0.119 & 0.116 \\  
13 & 0.141 & 0.123 & 0.115 & 0.116 \\  
14 & 0.068 & 0.097 & 0.107 & 0.109 \\  
15 & 0.045 & 0.09 & 0.105 & 0.105 \\  
16 & 0.057 & 0.12 & 0.11 & 0.12 \\  
17 & 0.101 & 0.18 & 0.135 & 0.142 \\  
18 & 0.165 & 0.236 & 0.165 & 0.165 \\  
19 & 0.239 & 0.219 & 0.193 & 0.166 \\  
20 & 0.304 & 0.257 & 0.224 & 0.167 \\  
21 & 0.358 & 0.306 & 0.227 & 0.174 \\  
22 & 0.403 & 0.322 & 0.223 & 0.181 \\  
23 & 0.424 & 0.328 & 0.223 & 0.176 \\  
24 & 0.423 & 0.302 & 0.22 & 0.159 \\  
25 & 0.407 & 0.278 & 0.227 & 0.165 \\  
26 & 0.333 & 0.26 & 0.201 & 0.154 \\  
27 & 0.279 & 0.226 & 0.175 & 0.141 \\  
28 & 0.232 & 0.18 & 0.183 & 0.142 \\  
29 & 0.191 & 0.166 & 0.163 & 0.141 \\  
30 & 0.162 & 0.165 & 0.159 & 0.145 \\  
31 & 0.181 & 0.16 & 0.15 & 0.132 \\  
32 & 0.254 & 0.189 & 0.152 & 0.123 \\  
33 & 0.323 & 0.238 & 0.159 & 0.115 \\  
34 & 0.387 & 0.254 & 0.184 & 0.142 \\  
35 & 0.462 & 0.27 & 0.192 & 0.135 \\  
36 & 0.505 & 0.266 & 0.192 & 0.14 \\  
37 & 0.499 & 0.244 & 0.17 & 0.139 \\  
38 & 0.449 & 0.209 & 0.139 & 0.131 \\  
39 & 0.341 & 0.179 & 0.126 & 0.124 \\  
40 & 0.257 & 0.144 & 0.129 & 0.127 \\  
41 & 0.192 & 0.136 & 0.131 & 0.126 \\  
42 & 0.123 & 0.127 & 0.126 & 0.119 \\  
43 & 0.093 & 0.13 & 0.133 & 0.12 \\  
44 & 0.079 & 0.132 & 0.137 & 0.124 \\  
45 & 0.105 & 0.144 & 0.143 & 0.125 \\  
46 & 0.157 & 0.156 & 0.133 & 0.119 \\  
47 & 0.22 & 0.164 & 0.152 & 0.121 \\  
48 & 0.273 & 0.186 & 0.143 & 0.123 \\  
49 & 0.281 & 0.206 & 0.154 & 0.125 \\  
50 & 0.286 & 0.197 & 0.137 & 0.122 \\  
51 & 0.283 & 0.166 & 0.119 & 0.116 \\  
52 & 0.256 & 0.165 & 0.121 & 0.116 \\  
53 & 0.23 & 0.164 & 0.126 & 0.126 \\  
54 & 0.196 & 0.184 & 0.14 & 0.127 \\  
55 & 0.221 & 0.225 & 0.164 & 0.138 \\  
56 & 0.258 & 0.262 & 0.177 & 0.133 \\  
57 & 0.331 & 0.289 & 0.187 & 0.138 \\  
58 & 0.421 & 0.264 & 0.166 & 0.151 \\  
59 & 0.521 & 0.263 & 0.179 & 0.149 \\  
60 & 0.588 & 0.273 & 0.195 & 0.141 \\  
61 & 0.678 & 0.305 & 0.234 & 0.145 \\  
62 & 0.692 & 0.327 & 0.239 & 0.164 \\  
63 & 0.667 & 0.3 & 0.218 & 0.18 \\  
64 & 0.581 & 0.279 & 0.181 & 0.171 \\  
65 & 0.478 & 0.235 & 0.165 & 0.167 \\  
66 & 0.399 & 0.197 & 0.127 & 0.147 \\  
67 & 0.29 & 0.163 & 0.118 & 0.113 \\  
68 & 0.213 & 0.136 & 0.117 & 0.106 \\  
69 & 0.149 & 0.133 & 0.111 & 0.115 \\  
70 & 0.093 & 0.119 & 0.118 & 0.109 \\  
71 & 0.058 & 0.105 & 0.105 & 0.114 \\  
72 & 0.056 & 0.086 & 0.109 & 0.115 \\  
73 & 0.087 & 0.087 & 0.116 & 0.122 \\  
74 & 0.147 & 0.117 & 0.123 & 0.15 \\  
75 & 0.227 & 0.163 & 0.123 & 0.155 \\  
76 & 0.366 & 0.226 & 0.14 & 0.143 \\  
77 & 0.487 & 0.278 & 0.164 & 0.156 \\  
78 & 0.586 & 0.346 & 0.194 & 0.171 \\  
79 & 0.605 & 0.282 & 0.185 & 0.14 \\  
80 & 0.622 & 0.293 & 0.184 & 0.153 \\  
81 & 0.59 & 0.299 & 0.192 & 0.146 \\  
82 & 0.552 & 0.317 & 0.141 & 0.143 \\  
83 & 0.4 & 0.242 & 0.135 & 0.127 \\  
84 & 0.238 & 0.174 & 0.119 & 0.121 \\  
85 & 0.113 & 0.115 & 0.113 & 0.103 \\  
86 & 0.045 & 0.089 & 0.098 & 0.12 \\  
87 & 0.042 & 0.09 & 0.118 & 0.111 \\  
88 & 0.116 & 0.132 & 0.15 & 0.123 \\  
89 & 0.2 & 0.219 & 0.167 & 0.122 \\  
90 & 0.304 & 0.288 & 0.221 & 0.138 \\  
91 & 0.437 & 0.321 & 0.23 & 0.144 \\  
92 & 0.586 & 0.379 & 0.254 & 0.157 \\  
93 & 0.686 & 0.418 & 0.276 & 0.161 \\  
94 & 0.749 & 0.455 & 0.257 & 0.172 \\  
95 & 0.739 & 0.439 & 0.236 & 0.175 \\  
\hline
\caption{The probability of measuring one in the ancilla qubit for the Hadamard test to determine the real part of $\bra{\psi} e^{-i H t} \ket{\psi}$ for $g = 0.5 $ after applying the measurement noise mitigation procedure described in Appendix \ref{appendix:mitigation} for different numbers of CNOT gates. Each entry in this table was calculated with $8192$ measurements.} 
\end{longtable} 
 
 \begin{longtable}{| c | c | c | c | c |} 
\hline 
\multicolumn{5}{|c|}{Error Mitigated Imaginary Hadamard Test P(0) g = 0.5} \\ 
\hline 
Time Slice & r=1 & r=3 & r=5 & r=7 \\ 
\hline 
1 & 0.224 & 0.111 & 0.099 & 0.105 \\  
2 & 0.126 & 0.082 & 0.105 & 0.093 \\  
3 & 0.058 & 0.074 & 0.105 & 0.103 \\  
4 & 0.041 & 0.101 & 0.12 & 0.114 \\  
5 & 0.071 & 0.129 & 0.127 & 0.124 \\  
6 & 0.168 & 0.17 & 0.143 & 0.143 \\  
7 & 0.301 & 0.225 & 0.19 & 0.158 \\  
8 & 0.409 & 0.292 & 0.213 & 0.152 \\  
9 & 0.536 & 0.341 & 0.221 & 0.164 \\  
10 & 0.629 & 0.374 & 0.233 & 0.167 \\  
11 & 0.663 & 0.373 & 0.225 & 0.13 \\  
12 & 0.644 & 0.333 & 0.196 & 0.131 \\  
13 & 0.567 & 0.328 & 0.195 & 0.144 \\  
14 & 0.456 & 0.277 & 0.16 & 0.12 \\  
15 & 0.367 & 0.207 & 0.149 & 0.112 \\  
16 & 0.264 & 0.139 & 0.145 & 0.113 \\  
17 & 0.196 & 0.089 & 0.102 & 0.092 \\  
18 & 0.107 & 0.069 & 0.109 & 0.101 \\  
19 & 0.074 & 0.075 & 0.099 & 0.105 \\  
20 & 0.052 & 0.085 & 0.109 & 0.111 \\  
21 & 0.064 & 0.107 & 0.129 & 0.126 \\  
22 & 0.093 & 0.148 & 0.169 & 0.153 \\  
23 & 0.109 & 0.185 & 0.171 & 0.182 \\  
24 & 0.16 & 0.24 & 0.192 & 0.192 \\  
25 & 0.2 & 0.235 & 0.173 & 0.164 \\  
26 & 0.223 & 0.27 & 0.171 & 0.176 \\  
27 & 0.246 & 0.239 & 0.176 & 0.158 \\  
28 & 0.251 & 0.229 & 0.202 & 0.151 \\  
29 & 0.232 & 0.208 & 0.182 & 0.169 \\  
30 & 0.22 & 0.158 & 0.177 & 0.148 \\  
31 & 0.176 & 0.153 & 0.154 & 0.144 \\  
32 & 0.153 & 0.126 & 0.15 & 0.135 \\  
33 & 0.152 & 0.137 & 0.143 & 0.129 \\  
34 & 0.148 & 0.177 & 0.168 & 0.15 \\  
35 & 0.241 & 0.19 & 0.18 & 0.155 \\  
36 & 0.349 & 0.242 & 0.198 & 0.147 \\  
37 & 0.469 & 0.262 & 0.219 & 0.159 \\  
38 & 0.556 & 0.305 & 0.222 & 0.152 \\  
39 & 0.604 & 0.303 & 0.195 & 0.14 \\  
40 & 0.541 & 0.246 & 0.157 & 0.137 \\  
41 & 0.498 & 0.228 & 0.157 & 0.129 \\  
42 & 0.449 & 0.206 & 0.147 & 0.128 \\  
43 & 0.386 & 0.176 & 0.143 & 0.127 \\  
44 & 0.296 & 0.169 & 0.129 & 0.12 \\  
45 & 0.208 & 0.14 & 0.127 & 0.116 \\  
46 & 0.129 & 0.124 & 0.123 & 0.128 \\  
47 & 0.09 & 0.112 & 0.13 & 0.105 \\  
48 & 0.09 & 0.115 & 0.123 & 0.113 \\  
49 & 0.113 & 0.126 & 0.132 & 0.117 \\  
50 & 0.159 & 0.127 & 0.137 & 0.114 \\  
51 & 0.182 & 0.134 & 0.135 & 0.112 \\  
52 & 0.219 & 0.115 & 0.12 & 0.11 \\  
53 & 0.204 & 0.123 & 0.135 & 0.108 \\  
54 & 0.166 & 0.127 & 0.122 & 0.112 \\  
55 & 0.166 & 0.12 & 0.128 & 0.121 \\  
56 & 0.13 & 0.139 & 0.137 & 0.112 \\  
57 & 0.11 & 0.149 & 0.137 & 0.124 \\  
58 & 0.122 & 0.14 & 0.134 & 0.137 \\  
59 & 0.171 & 0.153 & 0.16 & 0.148 \\  
60 & 0.259 & 0.186 & 0.166 & 0.15 \\  
61 & 0.339 & 0.243 & 0.195 & 0.178 \\  
62 & 0.44 & 0.302 & 0.231 & 0.203 \\  
63 & 0.537 & 0.379 & 0.257 & 0.215 \\  
64 & 0.601 & 0.411 & 0.263 & 0.208 \\  
65 & 0.638 & 0.425 & 0.268 & 0.206 \\  
66 & 0.665 & 0.402 & 0.238 & 0.19 \\  
67 & 0.639 & 0.316 & 0.206 & 0.132 \\  
68 & 0.611 & 0.315 & 0.202 & 0.139 \\  
69 & 0.546 & 0.311 & 0.205 & 0.14 \\  
70 & 0.454 & 0.271 & 0.192 & 0.121 \\  
71 & 0.358 & 0.242 & 0.16 & 0.118 \\  
72 & 0.27 & 0.195 & 0.136 & 0.1 \\  
73 & 0.175 & 0.162 & 0.148 & 0.109 \\  
74 & 0.09 & 0.123 & 0.126 & 0.109 \\  
75 & 0.053 & 0.104 & 0.122 & 0.103 \\  
76 & 0.042 & 0.085 & 0.113 & 0.104 \\  
77 & 0.075 & 0.098 & 0.112 & 0.109 \\  
78 & 0.134 & 0.124 & 0.11 & 0.12 \\  
79 & 0.229 & 0.152 & 0.125 & 0.118 \\  
80 & 0.34 & 0.195 & 0.148 & 0.123 \\  
81 & 0.469 & 0.238 & 0.175 & 0.137 \\  
82 & 0.657 & 0.398 & 0.236 & 0.166 \\  
83 & 0.687 & 0.393 & 0.231 & 0.161 \\  
84 & 0.653 & 0.378 & 0.216 & 0.145 \\  
85 & 0.569 & 0.372 & 0.198 & 0.143 \\  
86 & 0.436 & 0.31 & 0.163 & 0.134 \\  
87 & 0.301 & 0.212 & 0.137 & 0.135 \\  
88 & 0.176 & 0.147 & 0.133 & 0.114 \\  
89 & 0.094 & 0.095 & 0.108 & 0.106 \\  
90 & 0.045 & 0.093 & 0.114 & 0.103 \\  
91 & 0.051 & 0.112 & 0.135 & 0.115 \\  
92 & 0.099 & 0.144 & 0.159 & 0.127 \\  
93 & 0.195 & 0.207 & 0.187 & 0.139 \\  
94 & 0.339 & 0.251 & 0.229 & 0.159 \\  
95 & 0.505 & 0.331 & 0.261 & 0.172 \\  
\hline 
\caption{The probability of measuring zero in the ancilla qubit for the Hadamard test to determine the imaginary part of $\bra{\psi} e^{-i H t} \ket{\psi}$ for $g = 0.5 $ after applying the measurement noise mitigation procedure described in Appendix \ref{appendix:mitigation} for different numbers of CNOT gates. Each entry in this table was calculated with $8192$ measurements.}
\end{longtable} 
 
 \begin{longtable}{| c | c | c | c | c |} 
\hline 
\multicolumn{5}{|c|}{Error Mitigated Imaginary Hadamard Test P(1) g = 0.5} \\ 
\hline 
Time Slice & r=1 & r=3 & r=5 & r=7 \\ 
\hline 
1 & 0.53 & 0.372 & 0.231 & 0.16 \\  
2 & 0.633 & 0.397 & 0.253 & 0.179 \\  
3 & 0.712 & 0.39 & 0.256 & 0.167 \\  
4 & 0.693 & 0.364 & 0.239 & 0.177 \\  
5 & 0.644 & 0.351 & 0.226 & 0.166 \\  
6 & 0.536 & 0.302 & 0.203 & 0.154 \\  
7 & 0.437 & 0.26 & 0.181 & 0.15 \\  
8 & 0.308 & 0.188 & 0.154 & 0.149 \\  
9 & 0.162 & 0.134 & 0.135 & 0.12 \\  
10 & 0.084 & 0.1 & 0.125 & 0.11 \\  
11 & 0.04 & 0.098 & 0.117 & 0.107 \\  
12 & 0.059 & 0.125 & 0.126 & 0.112 \\  
13 & 0.125 & 0.141 & 0.14 & 0.137 \\  
14 & 0.238 & 0.181 & 0.182 & 0.135 \\  
15 & 0.354 & 0.237 & 0.202 & 0.146 \\  
16 & 0.454 & 0.349 & 0.245 & 0.185 \\  
17 & 0.531 & 0.426 & 0.268 & 0.213 \\  
18 & 0.604 & 0.431 & 0.261 & 0.197 \\  
19 & 0.622 & 0.396 & 0.234 & 0.176 \\  
20 & 0.613 & 0.374 & 0.205 & 0.176 \\  
21 & 0.579 & 0.333 & 0.186 & 0.16 \\  
22 & 0.502 & 0.285 & 0.157 & 0.136 \\  
23 & 0.452 & 0.226 & 0.15 & 0.111 \\  
24 & 0.369 & 0.177 & 0.144 & 0.104 \\  
25 & 0.28 & 0.153 & 0.145 & 0.117 \\  
26 & 0.218 & 0.126 & 0.127 & 0.115 \\  
27 & 0.177 & 0.129 & 0.129 & 0.108 \\  
28 & 0.157 & 0.123 & 0.112 & 0.12 \\  
29 & 0.188 & 0.134 & 0.12 & 0.117 \\  
30 & 0.224 & 0.181 & 0.126 & 0.126 \\  
31 & 0.28 & 0.196 & 0.136 & 0.117 \\  
32 & 0.331 & 0.221 & 0.144 & 0.115 \\  
33 & 0.355 & 0.226 & 0.145 & 0.124 \\  
34 & 0.371 & 0.199 & 0.143 & 0.122 \\  
35 & 0.318 & 0.202 & 0.133 & 0.119 \\  
36 & 0.239 & 0.156 & 0.117 & 0.117 \\  
37 & 0.151 & 0.13 & 0.121 & 0.12 \\  
38 & 0.099 & 0.118 & 0.13 & 0.126 \\  
39 & 0.057 & 0.122 & 0.145 & 0.132 \\  
40 & 0.083 & 0.145 & 0.142 & 0.129 \\  
41 & 0.128 & 0.181 & 0.163 & 0.13 \\  
42 & 0.197 & 0.201 & 0.176 & 0.129 \\  
43 & 0.265 & 0.222 & 0.165 & 0.14 \\  
44 & 0.331 & 0.229 & 0.179 & 0.133 \\  
45 & 0.4 & 0.244 & 0.177 & 0.137 \\  
46 & 0.464 & 0.242 & 0.174 & 0.13 \\  
47 & 0.479 & 0.244 & 0.162 & 0.127 \\  
48 & 0.433 & 0.239 & 0.157 & 0.134 \\  
49 & 0.354 & 0.218 & 0.137 & 0.122 \\  
50 & 0.275 & 0.186 & 0.13 & 0.118 \\  
51 & 0.241 & 0.179 & 0.111 & 0.12 \\  
52 & 0.222 & 0.166 & 0.126 & 0.13 \\  
53 & 0.239 & 0.178 & 0.128 & 0.129 \\  
54 & 0.259 & 0.203 & 0.142 & 0.131 \\  
55 & 0.332 & 0.237 & 0.155 & 0.134 \\  
56 & 0.39 & 0.25 & 0.15 & 0.134 \\  
57 & 0.444 & 0.241 & 0.15 & 0.124 \\  
58 & 0.47 & 0.23 & 0.129 & 0.126 \\  
59 & 0.47 & 0.205 & 0.13 & 0.123 \\  
60 & 0.439 & 0.193 & 0.135 & 0.134 \\  
61 & 0.385 & 0.159 & 0.141 & 0.107 \\  
62 & 0.311 & 0.127 & 0.117 & 0.1 \\  
63 & 0.206 & 0.098 & 0.105 & 0.105 \\  
64 & 0.121 & 0.086 & 0.091 & 0.104 \\  
65 & 0.065 & 0.081 & 0.094 & 0.104 \\  
66 & 0.04 & 0.087 & 0.107 & 0.105 \\  
67 & 0.07 & 0.107 & 0.118 & 0.115 \\  
68 & 0.086 & 0.11 & 0.14 & 0.128 \\  
69 & 0.125 & 0.118 & 0.143 & 0.116 \\  
70 & 0.182 & 0.131 & 0.132 & 0.113 \\  
71 & 0.269 & 0.14 & 0.14 & 0.122 \\  
72 & 0.351 & 0.191 & 0.149 & 0.134 \\  
73 & 0.424 & 0.241 & 0.174 & 0.172 \\  
74 & 0.514 & 0.291 & 0.207 & 0.182 \\  
75 & 0.564 & 0.341 & 0.209 & 0.198 \\  
76 & 0.601 & 0.371 & 0.22 & 0.197 \\  
77 & 0.624 & 0.369 & 0.222 & 0.189 \\  
78 & 0.59 & 0.326 & 0.225 & 0.172 \\  
79 & 0.439 & 0.238 & 0.176 & 0.139 \\  
80 & 0.337 & 0.192 & 0.153 & 0.126 \\  
81 & 0.207 & 0.144 & 0.119 & 0.126 \\  
82 & 0.101 & 0.107 & 0.113 & 0.119 \\  
83 & 0.059 & 0.096 & 0.132 & 0.121 \\  
84 & 0.082 & 0.119 & 0.153 & 0.122 \\  
85 & 0.169 & 0.15 & 0.164 & 0.133 \\  
86 & 0.311 & 0.223 & 0.202 & 0.145 \\  
87 & 0.451 & 0.289 & 0.224 & 0.144 \\  
88 & 0.576 & 0.376 & 0.252 & 0.16 \\  
89 & 0.672 & 0.422 & 0.284 & 0.173 \\  
90 & 0.722 & 0.435 & 0.295 & 0.183 \\  
91 & 0.721 & 0.388 & 0.259 & 0.15 \\  
92 & 0.679 & 0.365 & 0.224 & 0.147 \\  
93 & 0.593 & 0.326 & 0.189 & 0.139 \\  
94 & 0.454 & 0.289 & 0.15 & 0.115 \\  
95 & 0.296 & 0.208 & 0.126 & 0.118 \\  
\hline 
\caption{The probability of measuring one in the ancilla qubit for the Hadamard test to determine the imaginary part of $\bra{\psi} e^{-i H t} \ket{\psi}$ for $g = 0.5 $ after applying the measurement noise mitigation procedure described in Appendix \ref{appendix:mitigation} for different numbers of CNOT gates. Each entry in this table was calculated with $8192$ measurements.}
\end{longtable} 
 
\begin{longtable}{| c | c | c | c | c |} 
\hline 
\multicolumn{5}{|c|}{Error Mitigated Real Hadamard Test P(0) g = 1.} \\ 
\hline 
Time Slice & r=1 & r=3 & r=5 & r=7 \\ 
\hline 
1 & 0.445 & 0.134 & 0.131 & 0.089 \\  
2 & 0.335 & 0.136 & 0.128 & 0.119 \\  
3 & 0.35 & 0.154 & 0.124 & 0.125 \\  
4 & 0.318 & 0.152 & 0.117 & 0.099 \\  
5 & 0.239 & 0.192 & 0.118 & 0.095 \\  
6 & 0.187 & 0.179 & 0.124 & 0.105 \\  
7 & 0.12 & 0.199 & 0.108 & 0.076 \\  
8 & 0.141 & 0.183 & 0.123 & 0.111 \\  
9 & 0.109 & 0.218 & 0.092 & 0.117 \\  
10 & 0.102 & 0.207 & 0.121 & 0.121 \\  
11 & 0.099 & 0.219 & 0.119 & 0.127 \\  
12 & 0.137 & 0.228 & 0.127 & 0.112 \\  
13 & 0.164 & 0.204 & 0.132 & 0.113 \\  
14 & 0.167 & 0.215 & 0.139 & 0.121 \\  
15 & 0.187 & 0.203 & 0.128 & 0.116 \\  
16 & 0.25 & 0.178 & 0.127 & 0.119 \\  
17 & 0.281 & 0.197 & 0.132 & 0.112 \\  
18 & 0.294 & 0.177 & 0.104 & 0.126 \\  
19 & 0.345 & 0.166 & 0.115 & 0.117 \\  
20 & 0.394 & 0.179 & 0.136 & 0.116 \\  
21 & 0.387 & 0.173 & 0.138 & 0.113 \\  
22 & 0.434 & 0.155 & 0.126 & 0.11 \\  
23 & 0.424 & 0.151 & 0.136 & 0.122 \\  
24 & 0.422 & 0.143 & 0.127 & 0.117 \\  
25 & 0.378 & 0.125 & 0.13 & 0.116 \\  
26 & 0.367 & 0.149 & 0.138 & 0.115 \\  
27 & 0.394 & 0.118 & 0.132 & 0.113 \\  
28 & 0.372 & 0.135 & 0.13 & 0.104 \\  
29 & 0.32 & 0.139 & 0.122 & 0.077 \\  
30 & 0.331 & 0.125 & 0.113 & 0.116 \\  
31 & 0.324 & 0.124 & 0.106 & 0.118 \\  
32 & 0.291 & 0.139 & 0.121 & 0.11 \\  
33 & 0.321 & 0.133 & 0.147 & 0.124 \\  
34 & 0.273 & 0.143 & 0.143 & 0.117 \\  
35 & 0.272 & 0.134 & 0.133 & 0.122 \\  
36 & 0.254 & 0.141 & 0.123 & 0.119 \\  
37 & 0.183 & 0.148 & 0.147 & 0.1 \\  
38 & 0.183 & 0.147 & 0.131 & 0.108 \\  
39 & 0.131 & 0.145 & 0.139 & 0.116 \\  
40 & 0.124 & 0.155 & 0.153 & 0.118 \\  
41 & 0.112 & 0.177 & 0.143 & 0.103 \\  
42 & 0.094 & 0.161 & 0.139 & 0.104 \\  
43 & 0.094 & 0.169 & 0.121 & 0.1 \\  
44 & 0.099 & 0.175 & 0.13 & 0.093 \\  
45 & 0.11 & 0.192 & 0.122 & 0.1 \\  
46 & 0.111 & 0.172 & 0.111 & 0.112 \\  
47 & 0.137 & 0.183 & 0.135 & 0.133 \\  
48 & 0.134 & 0.192 & 0.116 & 0.098 \\  
49 & 0.175 & 0.158 & 0.109 & 0.095 \\  
50 & 0.189 & 0.16 & 0.122 & 0.117 \\  
51 & 0.174 & 0.156 & 0.129 & 0.108 \\  
52 & 0.141 & 0.148 & 0.106 & 0.113 \\  
53 & 0.169 & 0.16 & 0.129 & 0.103 \\  
54 & 0.162 & 0.157 & 0.122 & 0.125 \\  
55 & 0.189 & 0.157 & 0.106 & 0.11 \\  
56 & 0.158 & 0.16 & 0.114 & 0.106 \\  
57 & 0.183 & 0.137 & 0.124 & 0.121 \\  
58 & 0.177 & 0.151 & 0.134 & 0.102 \\  
59 & 0.192 & 0.164 & 0.122 & 0.083 \\  
60 & 0.229 & 0.179 & 0.101 & 0.119 \\  
61 & 0.232 & 0.168 & 0.112 & 0.115 \\  
62 & 0.279 & 0.154 & 0.121 & 0.102 \\  
63 & 0.282 & 0.154 & 0.123 & 0.097 \\  
64 & 0.32 & 0.13 & 0.103 & 0.093 \\  
65 & 0.368 & 0.168 & 0.119 & 0.104 \\  
66 & 0.392 & 0.144 & 0.112 & 0.101 \\  
67 & 0.379 & 0.145 & 0.113 & 0.108 \\  
68 & 0.417 & 0.111 & 0.105 & 0.109 \\  
69 & 0.409 & 0.121 & 0.117 & 0.103 \\  
70 & 0.417 & 0.119 & 0.117 & 0.103 \\  
71 & 0.364 & 0.11 & 0.127 & 0.105 \\  
72 & 0.366 & 0.099 & 0.117 & 0.099 \\  
73 & 0.338 & 0.119 & 0.123 & 0.122 \\  
74 & 0.288 & 0.108 & 0.121 & 0.099 \\  
75 & 0.247 & 0.119 & 0.111 & 0.115 \\  
76 & 0.23 & 0.122 & 0.09 & 0.104 \\  
77 & 0.152 & 0.122 & 0.101 & 0.113 \\  
78 & 0.137 & 0.136 & 0.111 & 0.1 \\  
79 & 0.092 & 0.129 & 0.115 & 0.117 \\  
80 & 0.07 & 0.127 & 0.083 & 0.075 \\  
81 & 0.063 & 0.106 & 0.088 & 0.079 \\  
82 & 0.109 & 0.154 & 0.1 & 0.083 \\  
83 & 0.073 & 0.163 & 0.066 & 0.079 \\  
84 & 0.111 & 0.11 & 0.098 & 0.097 \\  
85 & 0.136 & 0.153 & 0.117 & 0.113 \\  
86 & 0.144 & 0.136 & 0.101 & 0.097 \\  
87 & 0.202 & 0.124 & 0.093 & 0.102 \\  
88 & 0.162 & 0.155 & 0.117 & 0.076 \\  
89 & 0.179 & 0.135 & 0.099 & 0.117 \\  
90 & 0.257 & 0.169 & 0.084 & 0.074 \\  
91 & 0.336 & 0.133 & 0.106 & 0.103 \\  
92 & 0.245 & 0.177 & 0.095 & 0.076 \\  
93 & 0.492 & 0.187 & 0.106 & 0.091 \\  
94 & 0.473 & 0.18 & 0.137 & 0.116 \\  
95 & 0.43 & 0.184 & 0.118 & 0.112 \\  
96 & 0.24 & 0.107 & 0.113 & 0.119 \\  
97 & 0.244 & 0.108 & 0.082 & 0.091 \\  
98 & 0.411 & 0.132 & 0.115 & 0.114 \\  
99 & 0.353 & 0.141 & 0.133 & 0.101 \\  
100 & 0.336 & 0.111 & 0.102 & 0.096 \\  
101 & 0.325 & 0.14 & 0.104 & 0.108 \\  
102 & 0.25 & 0.109 & 0.101 & 0.124 \\  
103 & 0.176 & 0.133 & 0.12 & 0.11 \\  
104 & 0.124 & 0.116 & 0.13 & 0.127 \\  
105 & 0.112 & 0.123 & 0.131 & 0.109 \\  
106 & 0.114 & 0.112 & 0.136 & 0.123 \\  
107 & 0.077 & 0.142 & 0.131 & 0.114 \\  
108 & 0.053 & 0.139 & 0.125 & 0.102 \\  
109 & 0.076 & 0.175 & 0.138 & 0.129 \\  
110 & 0.09 & 0.203 & 0.147 & 0.116 \\  
111 & 0.11 & 0.19 & 0.112 & 0.113 \\  
112 & 0.173 & 0.233 & 0.136 & 0.121 \\  
\hline 
\caption{The probability of measuring zero in the ancilla qubit for the Hadamard test to determine the real part of $\bra{\psi} e^{-i H t} \ket{\psi}$ for $g = 1 $ after applying the measurement noise mitigation procedure described in Appendix \ref{appendix:mitigation} for different numbers of CNOT gates. Each entry in this table was calculated with $8192$ measurements.}
\end{longtable} 
 
 \begin{longtable}{| c | c | c | c | c |} 
\hline 
\multicolumn{5}{|c|}{Error Mitigated Real Hadamard Test P(1) g = 1.} \\ 
\hline 
Time Slice & r=1 & r=3 & r=5 & r=7 \\ 
\hline 
1 & 0.174 & 0.2 & 0.123 & 0.119 \\  
2 & 0.204 & 0.192 & 0.125 & 0.094 \\  
3 & 0.286 & 0.197 & 0.114 & 0.12 \\  
4 & 0.36 & 0.184 & 0.134 & 0.115 \\  
5 & 0.416 & 0.189 & 0.098 & 0.117 \\  
6 & 0.451 & 0.17 & 0.119 & 0.119 \\  
7 & 0.489 & 0.171 & 0.144 & 0.13 \\  
8 & 0.491 & 0.169 & 0.115 & 0.114 \\  
9 & 0.496 & 0.139 & 0.12 & 0.112 \\  
10 & 0.51 & 0.145 & 0.127 & 0.097 \\  
11 & 0.507 & 0.119 & 0.109 & 0.112 \\  
12 & 0.45 & 0.119 & 0.104 & 0.123 \\  
13 & 0.419 & 0.124 & 0.121 & 0.117 \\  
14 & 0.407 & 0.106 & 0.11 & 0.103 \\  
15 & 0.351 & 0.131 & 0.127 & 0.097 \\  
16 & 0.313 & 0.141 & 0.124 & 0.118 \\  
17 & 0.264 & 0.122 & 0.117 & 0.116 \\  
18 & 0.213 & 0.127 & 0.128 & 0.113 \\  
19 & 0.181 & 0.147 & 0.124 & 0.13 \\  
20 & 0.166 & 0.132 & 0.112 & 0.119 \\  
21 & 0.117 & 0.144 & 0.125 & 0.125 \\  
22 & 0.088 & 0.145 & 0.126 & 0.133 \\  
23 & 0.091 & 0.166 & 0.116 & 0.123 \\  
24 & 0.09 & 0.154 & 0.132 & 0.113 \\  
25 & 0.092 & 0.166 & 0.132 & 0.111 \\  
26 & 0.108 & 0.181 & 0.123 & 0.133 \\  
27 & 0.093 & 0.173 & 0.13 & 0.11 \\  
28 & 0.16 & 0.149 & 0.124 & 0.09 \\  
29 & 0.124 & 0.166 & 0.131 & 0.121 \\  
30 & 0.13 & 0.165 & 0.129 & 0.12 \\  
31 & 0.151 & 0.167 & 0.128 & 0.103 \\  
32 & 0.193 & 0.168 & 0.129 & 0.135 \\  
33 & 0.284 & 0.208 & 0.138 & 0.132 \\  
34 & 0.296 & 0.223 & 0.105 & 0.108 \\  
35 & 0.325 & 0.222 & 0.137 & 0.11 \\  
36 & 0.361 & 0.207 & 0.148 & 0.099 \\  
37 & 0.425 & 0.211 & 0.121 & 0.128 \\  
38 & 0.41 & 0.214 & 0.122 & 0.121 \\  
39 & 0.466 & 0.198 & 0.116 & 0.124 \\  
40 & 0.488 & 0.227 & 0.129 & 0.12 \\  
41 & 0.466 & 0.152 & 0.132 & 0.099 \\  
42 & 0.467 & 0.176 & 0.108 & 0.128 \\  
43 & 0.465 & 0.175 & 0.124 & 0.13 \\  
44 & 0.457 & 0.172 & 0.126 & 0.132 \\  
45 & 0.392 & 0.132 & 0.12 & 0.125 \\  
46 & 0.413 & 0.131 & 0.126 & 0.113 \\  
47 & 0.412 & 0.136 & 0.119 & 0.096 \\  
48 & 0.369 & 0.141 & 0.132 & 0.111 \\  
49 & 0.289 & 0.146 & 0.126 & 0.104 \\  
50 & 0.273 & 0.126 & 0.116 & 0.104 \\  
51 & 0.286 & 0.121 & 0.118 & 0.114 \\  
52 & 0.284 & 0.145 & 0.134 & 0.127 \\  
53 & 0.252 & 0.116 & 0.13 & 0.127 \\  
54 & 0.26 & 0.123 & 0.122 & 0.121 \\  
55 & 0.242 & 0.129 & 0.131 & 0.12 \\  
56 & 0.253 & 0.127 & 0.127 & 0.13 \\  
57 & 0.247 & 0.152 & 0.115 & 0.11 \\  
58 & 0.279 & 0.145 & 0.103 & 0.129 \\  
59 & 0.239 & 0.141 & 0.103 & 0.132 \\  
60 & 0.24 & 0.131 & 0.131 & 0.102 \\  
61 & 0.24 & 0.141 & 0.129 & 0.128 \\  
62 & 0.212 & 0.142 & 0.105 & 0.135 \\  
63 & 0.201 & 0.146 & 0.121 & 0.118 \\  
64 & 0.178 & 0.149 & 0.132 & 0.121 \\  
65 & 0.163 & 0.11 & 0.124 & 0.109 \\  
66 & 0.122 & 0.149 & 0.142 & 0.134 \\  
67 & 0.135 & 0.152 & 0.122 & 0.106 \\  
68 & 0.089 & 0.139 & 0.108 & 0.123 \\  
69 & 0.114 & 0.142 & 0.119 & 0.121 \\  
70 & 0.125 & 0.148 & 0.129 & 0.107 \\  
71 & 0.149 & 0.151 & 0.12 & 0.117 \\  
72 & 0.182 & 0.166 & 0.107 & 0.128 \\  
73 & 0.198 & 0.171 & 0.107 & 0.112 \\  
74 & 0.255 & 0.176 & 0.12 & 0.125 \\  
75 & 0.294 & 0.151 & 0.124 & 0.147 \\  
76 & 0.382 & 0.167 & 0.127 & 0.101 \\  
77 & 0.38 & 0.167 & 0.111 & 0.116 \\  
78 & 0.416 & 0.175 & 0.128 & 0.12 \\  
79 & 0.403 & 0.16 & 0.129 & 0.105 \\  
80 & 0.257 & 0.183 & 0.124 & 0.105 \\  
81 & 0.382 & 0.143 & 0.084 & 0.085 \\  
82 & 0.381 & 0.157 & 0.136 & 0.09 \\  
83 & 0.318 & 0.155 & 0.102 & 0.099 \\  
84 & 0.38 & 0.124 & 0.108 & 0.093 \\  
85 & 0.332 & 0.103 & 0.12 & 0.103 \\  
86 & 0.291 & 0.109 & 0.101 & 0.143 \\  
87 & 0.256 & 0.133 & 0.09 & 0.115 \\  
88 & 0.184 & 0.155 & 0.13 & 0.093 \\  
89 & 0.123 & 0.116 & 0.108 & 0.095 \\  
90 & 0.132 & 0.118 & 0.089 & 0.09 \\  
91 & 0.198 & 0.108 & 0.129 & 0.128 \\  
92 & 0.116 & 0.139 & 0.085 & 0.11 \\  
93 & 0.108 & 0.154 & 0.117 & 0.136 \\  
94 & 0.09 & 0.154 & 0.117 & 0.124 \\  
95 & 0.096 & 0.161 & 0.106 & 0.124 \\  
96 & 0.094 & 0.161 & 0.098 & 0.087 \\  
97 & 0.099 & 0.107 & 0.106 & 0.089 \\  
98 & 0.164 & 0.171 & 0.131 & 0.112 \\  
99 & 0.192 & 0.203 & 0.13 & 0.115 \\  
100 & 0.239 & 0.213 & 0.121 & 0.107 \\  
101 & 0.323 & 0.186 & 0.14 & 0.122 \\  
102 & 0.326 & 0.212 & 0.151 & 0.133 \\  
103 & 0.399 & 0.209 & 0.131 & 0.134 \\  
104 & 0.423 & 0.239 & 0.143 & 0.123 \\  
105 & 0.437 & 0.214 & 0.128 & 0.131 \\  
106 & 0.472 & 0.221 & 0.142 & 0.125 \\  
107 & 0.449 & 0.201 & 0.133 & 0.113 \\  
108 & 0.505 & 0.162 & 0.125 & 0.14 \\  
109 & 0.472 & 0.189 & 0.111 & 0.131 \\  
110 & 0.524 & 0.147 & 0.132 & 0.126 \\  
111 & 0.509 & 0.161 & 0.131 & 0.124 \\  
112 & 0.476 & 0.149 & 0.105 & 0.124 \\  
\hline 
\caption{The probability of measuring one in the ancilla qubit for the Hadamard test to determine the real part of $\bra{\psi} e^{-i H t} \ket{\psi}$ for $g = 1 $ after applying the measurement noise mitigation procedure described in Appendix \ref{appendix:mitigation} for different numbers of CNOT gates. Each entry in this table was calculated with $8192$ measurements.}
\end{longtable} 
 
 \begin{longtable}{| c | c | c | c | c |} 
\hline 
\multicolumn{5}{|c|}{Error Mitigated Imaginary Hadamard Test P(0) g = 1.} \\ 
\hline 
Time Slice & r=1 & r=3 & r=5 & r=7 \\ 
\hline 
1 & 0.125 & 0.178 & 0.112 & 0.125 \\  
2 & 0.037 & 0.194 & 0.134 & 0.088 \\  
3 & 0.104 & 0.225 & 0.125 & 0.12 \\  
4 & 0.099 & 0.195 & 0.112 & 0.105 \\  
5 & 0.112 & 0.208 & 0.134 & 0.106 \\  
6 & 0.145 & 0.188 & 0.121 & 0.095 \\  
7 & 0.197 & 0.228 & 0.127 & 0.086 \\  
8 & 0.21 & 0.224 & 0.11 & 0.1 \\  
9 & 0.275 & 0.207 & 0.104 & 0.116 \\  
10 & 0.317 & 0.205 & 0.118 & 0.103 \\  
11 & 0.36 & 0.217 & 0.127 & 0.11 \\  
12 & 0.426 & 0.209 & 0.122 & 0.104 \\  
13 & 0.409 & 0.181 & 0.127 & 0.113 \\  
14 & 0.462 & 0.167 & 0.116 & 0.112 \\  
15 & 0.45 & 0.168 & 0.096 & 0.127 \\  
16 & 0.457 & 0.155 & 0.105 & 0.12 \\  
17 & 0.457 & 0.137 & 0.117 & 0.1 \\  
18 & 0.425 & 0.147 & 0.119 & 0.104 \\  
19 & 0.409 & 0.12 & 0.105 & 0.116 \\  
20 & 0.397 & 0.163 & 0.122 & 0.104 \\  
21 & 0.371 & 0.122 & 0.118 & 0.113 \\  
22 & 0.377 & 0.14 & 0.116 & 0.12 \\  
23 & 0.318 & 0.138 & 0.125 & 0.12 \\  
24 & 0.295 & 0.123 & 0.13 & 0.112 \\  
25 & 0.226 & 0.126 & 0.123 & 0.113 \\  
26 & 0.253 & 0.121 & 0.127 & 0.101 \\  
27 & 0.178 & 0.136 & 0.148 & 0.095 \\  
28 & 0.155 & 0.125 & 0.121 & 0.108 \\  
29 & 0.129 & 0.125 & 0.126 & 0.108 \\  
30 & 0.123 & 0.122 & 0.136 & 0.09 \\  
31 & 0.108 & 0.139 & 0.129 & 0.122 \\  
32 & 0.082 & 0.124 & 0.127 & 0.112 \\  
33 & 0.08 & 0.173 & 0.122 & 0.122 \\  
34 & 0.074 & 0.176 & 0.142 & 0.119 \\  
35 & 0.083 & 0.227 & 0.15 & 0.122 \\  
36 & 0.085 & 0.196 & 0.142 & 0.123 \\  
37 & 0.131 & 0.227 & 0.131 & 0.11 \\  
38 & 0.148 & 0.205 & 0.125 & 0.124 \\  
39 & 0.151 & 0.187 & 0.123 & 0.113 \\  
40 & 0.178 & 0.224 & 0.144 & 0.127 \\  
41 & 0.197 & 0.201 & 0.113 & 0.097 \\  
42 & 0.226 & 0.206 & 0.118 & 0.111 \\  
43 & 0.284 & 0.211 & 0.113 & 0.108 \\  
44 & 0.308 & 0.205 & 0.12 & 0.108 \\  
45 & 0.338 & 0.211 & 0.134 & 0.115 \\  
46 & 0.338 & 0.174 & 0.112 & 0.118 \\  
47 & 0.378 & 0.174 & 0.118 & 0.138 \\  
48 & 0.378 & 0.142 & 0.126 & 0.118 \\  
49 & 0.328 & 0.156 & 0.115 & 0.107 \\  
50 & 0.335 & 0.164 & 0.126 & 0.117 \\  
51 & 0.318 & 0.167 & 0.124 & 0.128 \\  
52 & 0.298 & 0.134 & 0.106 & 0.121 \\  
53 & 0.315 & 0.139 & 0.112 & 0.119 \\  
54 & 0.289 & 0.146 & 0.123 & 0.112 \\  
55 & 0.284 & 0.152 & 0.117 & 0.11 \\  
56 & 0.287 & 0.15 & 0.109 & 0.117 \\  
57 & 0.31 & 0.149 & 0.114 & 0.123 \\  
58 & 0.321 & 0.15 & 0.138 & 0.118 \\  
59 & 0.313 & 0.134 & 0.13 & 0.12 \\  
60 & 0.328 & 0.164 & 0.111 & 0.118 \\  
61 & 0.364 & 0.139 & 0.124 & 0.122 \\  
62 & 0.372 & 0.131 & 0.12 & 0.128 \\  
63 & 0.358 & 0.129 & 0.132 & 0.096 \\  
64 & 0.363 & 0.123 & 0.11 & 0.114 \\  
65 & 0.382 & 0.124 & 0.127 & 0.121 \\  
66 & 0.331 & 0.127 & 0.12 & 0.103 \\  
67 & 0.286 & 0.114 & 0.126 & 0.107 \\  
68 & 0.271 & 0.108 & 0.103 & 0.105 \\  
69 & 0.234 & 0.103 & 0.102 & 0.109 \\  
70 & 0.196 & 0.125 & 0.113 & 0.103 \\  
71 & 0.18 & 0.112 & 0.111 & 0.107 \\  
72 & 0.116 & 0.126 & 0.105 & 0.125 \\  
73 & 0.109 & 0.118 & 0.11 & 0.125 \\  
74 & 0.099 & 0.118 & 0.117 & 0.124 \\  
75 & 0.075 & 0.135 & 0.113 & 0.114 \\  
76 & 0.084 & 0.156 & 0.132 & 0.114 \\  
77 & 0.103 & 0.167 & 0.123 & 0.125 \\  
78 & 0.143 & 0.202 & 0.099 & 0.098 \\  
79 & 0.153 & 0.153 & 0.101 & 0.116 \\  
80 & 0.146 & 0.109 & 0.096 & 0.069 \\  
81 & 0.216 & 0.172 & 0.105 & 0.087 \\  
82 & 0.298 & 0.178 & 0.101 & 0.112 \\  
83 & 0.248 & 0.151 & 0.102 & 0.092 \\  
84 & 0.369 & 0.125 & 0.086 & 0.123 \\  
85 & 0.366 & 0.154 & 0.107 & 0.084 \\  
86 & 0.317 & 0.124 & 0.085 & 0.104 \\  
87 & 0.348 & 0.105 & 0.107 & 0.131 \\  
88 & 0.261 & 0.156 & 0.092 & 0.089 \\  
89 & 0.232 & 0.119 & 0.106 & 0.1 \\  
90 & 0.284 & 0.144 & 0.099 & 0.082 \\  
91 & 0.377 & 0.119 & 0.108 & 0.11 \\  
92 & 0.227 & 0.153 & 0.101 & 0.091 \\  
93 & 0.391 & 0.147 & 0.114 & 0.119 \\  
94 & 0.35 & 0.13 & 0.126 & 0.112 \\  
95 & 0.258 & 0.136 & 0.111 & 0.114 \\  
96 & 0.118 & 0.116 & 0.102 & 0.112 \\  
97 & 0.121 & 0.108 & 0.073 & 0.111 \\  
98 & 0.138 & 0.13 & 0.128 & 0.091 \\  
99 & 0.123 & 0.132 & 0.122 & 0.104 \\  
100 & 0.063 & 0.124 & 0.113 & 0.115 \\  
101 & 0.092 & 0.153 & 0.126 & 0.108 \\  
102 & 0.08 & 0.162 & 0.142 & 0.125 \\  
103 & 0.093 & 0.203 & 0.15 & 0.117 \\  
104 & 0.107 & 0.176 & 0.136 & 0.134 \\  
105 & 0.161 & 0.212 & 0.133 & 0.108 \\  
106 & 0.144 & 0.188 & 0.154 & 0.122 \\  
107 & 0.195 & 0.208 & 0.126 & 0.137 \\  
108 & 0.228 & 0.193 & 0.148 & 0.114 \\  
109 & 0.293 & 0.238 & 0.143 & 0.133 \\  
110 & 0.372 & 0.212 & 0.123 & 0.104 \\  
111 & 0.42 & 0.24 & 0.144 & 0.121 \\  
112 & 0.513 & 0.233 & 0.139 & 0.11 \\  
\hline 
\caption{The probability of measuring zero in the ancilla qubit for the Hadamard test to determine the imaginary part of $\bra{\psi} e^{-i H t} \ket{\psi}$ for $g = 1 $ after applying the measurement noise mitigation procedure described in Appendix \ref{appendix:mitigation} for different numbers of CNOT gates. Each entry in this table was calculated with $8192$ measurements.}
\end{longtable} 
 
 \begin{longtable}{| c | c | c | c | c |} 
\hline 
\multicolumn{5}{|c|}{Error Mitigated Imaginary Hadamard Test P(1) g = 1.} \\ 
\hline 
Time Slice & r=1 & r=3 & r=5 & r=7 \\ 
\hline 
1 & 0.51 & 0.144 & 0.119 & 0.13 \\  
2 & 0.467 & 0.125 & 0.116 & 0.114 \\  
3 & 0.528 & 0.134 & 0.124 & 0.125 \\  
4 & 0.575 & 0.143 & 0.122 & 0.119 \\  
5 & 0.536 & 0.146 & 0.11 & 0.117 \\  
6 & 0.501 & 0.135 & 0.134 & 0.13 \\  
7 & 0.424 & 0.128 & 0.117 & 0.117 \\  
8 & 0.411 & 0.137 & 0.121 & 0.13 \\  
9 & 0.343 & 0.152 & 0.119 & 0.113 \\  
10 & 0.293 & 0.15 & 0.132 & 0.118 \\  
11 & 0.233 & 0.132 & 0.122 & 0.126 \\  
12 & 0.163 & 0.157 & 0.135 & 0.128 \\  
13 & 0.163 & 0.175 & 0.12 & 0.1 \\  
14 & 0.122 & 0.157 & 0.137 & 0.113 \\  
15 & 0.104 & 0.148 & 0.142 & 0.109 \\  
16 & 0.105 & 0.191 & 0.134 & 0.114 \\  
17 & 0.079 & 0.172 & 0.125 & 0.115 \\  
18 & 0.105 & 0.182 & 0.131 & 0.126 \\  
19 & 0.104 & 0.185 & 0.129 & 0.086 \\  
20 & 0.17 & 0.152 & 0.125 & 0.13 \\  
21 & 0.153 & 0.186 & 0.132 & 0.134 \\  
22 & 0.159 & 0.177 & 0.132 & 0.125 \\  
23 & 0.209 & 0.179 & 0.118 & 0.102 \\  
24 & 0.212 & 0.179 & 0.114 & 0.113 \\  
25 & 0.249 & 0.18 & 0.131 & 0.125 \\  
26 & 0.256 & 0.196 & 0.108 & 0.13 \\  
27 & 0.32 & 0.164 & 0.112 & 0.115 \\  
28 & 0.345 & 0.157 & 0.127 & 0.11 \\  
29 & 0.306 & 0.17 & 0.117 & 0.112 \\  
30 & 0.33 & 0.168 & 0.127 & 0.138 \\  
31 & 0.386 & 0.166 & 0.131 & 0.099 \\  
32 & 0.424 & 0.151 & 0.132 & 0.122 \\  
33 & 0.506 & 0.168 & 0.135 & 0.142 \\  
34 & 0.506 & 0.186 & 0.121 & 0.112 \\  
35 & 0.534 & 0.133 & 0.114 & 0.13 \\  
36 & 0.505 & 0.146 & 0.141 & 0.109 \\  
37 & 0.482 & 0.145 & 0.14 & 0.13 \\  
38 & 0.448 & 0.146 & 0.129 & 0.12 \\  
39 & 0.45 & 0.155 & 0.131 & 0.121 \\  
40 & 0.437 & 0.132 & 0.135 & 0.126 \\  
41 & 0.38 & 0.133 & 0.129 & 0.124 \\  
42 & 0.336 & 0.122 & 0.126 & 0.127 \\  
43 & 0.285 & 0.139 & 0.134 & 0.119 \\  
44 & 0.25 & 0.125 & 0.135 & 0.116 \\  
45 & 0.181 & 0.122 & 0.1 & 0.118 \\  
46 & 0.198 & 0.151 & 0.145 & 0.111 \\  
47 & 0.154 & 0.158 & 0.14 & 0.111 \\  
48 & 0.133 & 0.161 & 0.125 & 0.115 \\  
49 & 0.142 & 0.153 & 0.14 & 0.114 \\  
50 & 0.113 & 0.144 & 0.122 & 0.126 \\  
51 & 0.137 & 0.132 & 0.12 & 0.104 \\  
52 & 0.114 & 0.137 & 0.137 & 0.115 \\  
53 & 0.119 & 0.151 & 0.129 & 0.086 \\  
54 & 0.136 & 0.111 & 0.125 & 0.115 \\  
55 & 0.145 & 0.151 & 0.126 & 0.103 \\  
56 & 0.125 & 0.154 & 0.14 & 0.107 \\  
57 & 0.126 & 0.141 & 0.138 & 0.103 \\  
58 & 0.141 & 0.154 & 0.096 & 0.12 \\  
59 & 0.108 & 0.147 & 0.102 & 0.111 \\  
60 & 0.106 & 0.135 & 0.118 & 0.106 \\  
61 & 0.118 & 0.156 & 0.126 & 0.12 \\  
62 & 0.117 & 0.175 & 0.1 & 0.122 \\  
63 & 0.134 & 0.175 & 0.108 & 0.122 \\  
64 & 0.129 & 0.161 & 0.115 & 0.102 \\  
65 & 0.155 & 0.148 & 0.11 & 0.112 \\  
66 & 0.175 & 0.169 & 0.107 & 0.124 \\  
67 & 0.223 & 0.178 & 0.111 & 0.104 \\  
68 & 0.223 & 0.159 & 0.121 & 0.126 \\  
69 & 0.281 & 0.169 & 0.118 & 0.117 \\  
70 & 0.328 & 0.153 & 0.12 & 0.137 \\  
71 & 0.34 & 0.167 & 0.128 & 0.117 \\  
72 & 0.412 & 0.139 & 0.125 & 0.106 \\  
73 & 0.417 & 0.169 & 0.117 & 0.109 \\  
74 & 0.432 & 0.163 & 0.11 & 0.101 \\  
75 & 0.478 & 0.145 & 0.119 & 0.123 \\  
76 & 0.488 & 0.139 & 0.123 & 0.124 \\  
77 & 0.453 & 0.139 & 0.097 & 0.096 \\  
78 & 0.421 & 0.115 & 0.125 & 0.112 \\  
79 & 0.317 & 0.117 & 0.141 & 0.095 \\  
80 & 0.165 & 0.156 & 0.122 & 0.084 \\  
81 & 0.231 & 0.125 & 0.098 & 0.083 \\  
82 & 0.198 & 0.158 & 0.136 & 0.092 \\  
83 & 0.137 & 0.122 & 0.103 & 0.087 \\  
84 & 0.184 & 0.134 & 0.101 & 0.084 \\  
85 & 0.133 & 0.109 & 0.131 & 0.094 \\  
86 & 0.098 & 0.124 & 0.096 & 0.129 \\  
87 & 0.081 & 0.106 & 0.078 & 0.094 \\  
88 & 0.063 & 0.123 & 0.133 & 0.092 \\  
89 & 0.063 & 0.13 & 0.103 & 0.076 \\  
90 & 0.1 & 0.128 & 0.085 & 0.083 \\  
91 & 0.166 & 0.124 & 0.12 & 0.092 \\  
92 & 0.124 & 0.132 & 0.124 & 0.113 \\  
93 & 0.201 & 0.198 & 0.136 & 0.118 \\  
94 & 0.214 & 0.226 & 0.119 & 0.111 \\  
95 & 0.27 & 0.209 & 0.134 & 0.106 \\  
96 & 0.218 & 0.154 & 0.098 & 0.104 \\  
97 & 0.244 & 0.127 & 0.1 & 0.082 \\  
98 & 0.437 & 0.173 & 0.114 & 0.108 \\  
99 & 0.436 & 0.191 & 0.123 & 0.11 \\  
100 & 0.498 & 0.191 & 0.108 & 0.089 \\  
101 & 0.546 & 0.157 & 0.11 & 0.121 \\  
102 & 0.473 & 0.156 & 0.114 & 0.123 \\  
103 & 0.475 & 0.164 & 0.112 & 0.123 \\  
104 & 0.448 & 0.18 & 0.13 & 0.128 \\  
105 & 0.395 & 0.122 & 0.126 & 0.112 \\  
106 & 0.426 & 0.15 & 0.096 & 0.122 \\  
107 & 0.339 & 0.141 & 0.131 & 0.101 \\  
108 & 0.302 & 0.109 & 0.126 & 0.124 \\  
109 & 0.242 & 0.144 & 0.117 & 0.122 \\  
110 & 0.234 & 0.137 & 0.13 & 0.128 \\  
111 & 0.174 & 0.113 & 0.131 & 0.1 \\  
112 & 0.137 & 0.122 & 0.125 & 0.134 \\  
\hline
\caption{The probability of measuring one in the ancilla qubit for the Hadamard test to determine the imaginary part of $\bra{\psi} e^{-i H t} \ket{\psi}$ for $g = 1 $ after applying the measurement noise mitigation procedure described in Appendix \ref{appendix:mitigation} for different numbers of CNOT gates. Each entry in this table was calculated with $8192$ measurements.} 
\end{longtable} 
 
\end{document}